\documentstyle[seceq,epsf]{ptptex}

\newcommand{\lsim}
{\mathrel{\mbox{\raisebox{-1.0ex}
    {$\stackrel{\displaystyle <}{\displaystyle \sim}$}}}}

\title{
Phase Transition in Dense QCD \\ with the Schwinger-Dyson Equation
}

\author{
Masayasu {\sc Harada}\footnote{E-mail: harada@eken.phys.nagoya-u.ac.jp} 
and Satoshi {\sc Takagi}\footnote{E-mail: satoshi@eken.phys.nagoya-u.ac.jp}
}

\inst{
Department of Physics, Nagoya University, Nagoya, 464-8602, Japan
}

\recdate{November 14, 2001}

\abst{
We investigate the phase structure of dense QCD
using the Schwinger-Dyson equation (SDE) with the improved ladder
approximation in the Landau gauge.
We use the gluon propagator with the electric mode corrected by 
Debye screening and the magnetic mode corrected by Landau damping.
We solve the coupled SDE for the Majorana masses of the quark and antiquark
(separately from the SDE for the Dirac mass)
in the low density and intermediate density regions.
In the low density region, both the SDEs for the Majorana masses and the
Dirac mass have nontrivial solutions that correspond to the color symmetry
breaking (CSB) vacuum and the chiral symmetry breaking ($\chi$SB)
vacuum.
Comparing the values of the effective potential in the two vacua, we show
that the phase transition from the $\chi$SB vacuum to the CSB vacuum
is of first order.
The resultant value of the critical chemical potential is about 210\,MeV,
which is smaller than that obtained from the SDE analysis 
including only the Dirac mass.
The momentum dependences of the Majorana masses of the quark and
antiquark are shown to be quite different, 
although the masses are of the same order.
}

\begin{document}

\maketitle

\section{Introduction}
\label{Introduction}
Dynamical chiral symmetry breaking is one of the most important 
features of QCD.
The QCD Lagrangian of the light quark sector has an approximate chiral
symmetry, and 
it is spontaneously broken by the strong interaction of QCD.
Several physical processes of light hadrons in the low-energy region
are governed by the chiral symmetry property.
In hot and/or dense matter, on the other hand,
the chiral condensate (quark-antiquark condensate) melts, and 
the chiral symmetry is restored
(for recent reviews, see, e.g., Ref.~\citen{restoration}).

In early days of the research of such systems, 
it was pointed~\cite{Barr,Bailin}
that there exists a diquark condensate (color superconducting condensate)
 in the high density region.
Recently, it was shown~\cite{Al}
by using the instanton induced four-Fermi interaction model
that the diquark condensate is on the order of $100$\,MeV.
The diquark condensate is not a color singlet, and thus breaks the
color gauge symmetry.
This phenomenon is called color superconductivity
(for recent reviews, see, e.g., Ref.~\citen{RK}).

There are many works on color superconductivity employing analysis based on 
the SDE (see, e.g., Refs.~\citen{Hong}
--\citen{Rober}).
Since the SDE analysis is valid for weak coupling,
most of the works concentrate on the high density region.
However, the density inside a neutron star is believed to be 
close to the phase transition point(see, e.g., Ref.~\citen{RK}),
where the competition between the chiral condensate and the
diquark condensate becomes important.

Competition between the chiral
condensate and the diquark condensate
has been studied with several approaches
(see, e.g., Refs.~\citen{Be,Shuryak,Rober}).
In Ref.~\citen{Rober} the SDE is converted into an algebraic equation
using the confining model gluon propagator, and
it was shown that the color symmetry breaking vacuum exists
as a solution of the SDE, even in the low density region,
 but it is less stable than the chiral symmetry breaking vacuum.
It is interesting to study whether such a false vacuum exists 
by fully solving the SDE.

In most analyses previously carried out with the SDE,
only one Majorana mass is included:
The Majorana mass for the quark is analyzed while that for
the antiquark is dropped, or the Majorana masses of the two are assumed 
to be equal.Thus, it would also be interesting to include 
two Majorana masses in solving the SDE.

In this paper, 
we study the phase transition in QCD in dense matter
using the SDE with the improved ladder approximation 
in the Landau gauge.
In a dense medium, although the electric gluons have Debye masses,
the magnetic gluons are not screened.
The unscreened magnetic gluons give a dominant contribution to the
formation of the Majorana mass gap.~\cite{Son,Hong,Schaf3}
In the high density region,
through the hard dense loop approximation,
the effect of the Debye screening is included
in the electric mode of the gluon propagator,
and the effect of Landau damping is in 
the magnetic mode, as shown in Ref.~\citen{Hong}.
Here we extrapolate the same form
to the intermediate density region,
and solve the coupled SDE for
the Majorana masses of the quark and antiquark
separately from the SDE for the Dirac mass.
The true vacuum is determined 
by comparing the values of the effective
potential at the solution.

The symmetry breaking pattern in the color superconducting phase
depends on the number of light quarks.
When the number of massless quarks is two
($N_f=2$), the dominant
diquark condensate is the singlet of 
$SU(2)_{L}\times{SU}(2)_{R}$,
and thus the 
chiral symmetry is not broken, while 
the color ${SU}(3)_{c}$ is broken down to its subgroup 
${SU}(2)_{c}$ (2SC phase).~\cite{Al}
For three massless quarks ($N_f=3$), on the other hand,
the dominant diquark condensate breaks the color ${SU}(3)_{c}$
and the chiral ${SU}(3)_{L}\times{SU}(3)_{R}$
symmetry together into their diagonal subgroup
${SU}(3)_{c+L+R}$ 
[color flavor locking(CFL) phase).~\cite{Wilczek}
The breaking pattern in real QCD depends on the strange quark 
mass~(see, e.g., Ref.~\citen{Alford}).
The above-mentioned two breaking patterns are understood as two limits of the
strange quark mass:
When the strange quark is heavy enough, the 2SC phase is realized.
In contrast for the massless strange quark, the CFL phase is
realized.

In the present analysis
we regard $u$ and $d$ quarks as massless, but
we consider the current mass of the $s$-quark
to be large enough to neglect the strange quark in 
the formation of the diquark condensate of $u$ and $d$ quarks;
we assume that the color superconductivity is realized in 2SC phase. 
Furthermore, we ignore the effect of the strange quark 
in the effective potential, since we are interested in the differences among 
the values of the effective
potential corresponding to the CSB vacuum, the $\chi$SB vacuum
and the trivial vacuum.
In the running coupling, however,
the effect of the strange quark is included, because its mass is 
not larger than the energy region in which the dominant part 
of the support of the effective potential lies.
In other words, we include the strange quark only as
a sea quark in the present analysis.

The Lagrangian of the $u$ and $d$ quark sector
in the present analysis 
is invariant under the chiral ${SU}(2)_{L}\times{SU}(2)_{R}$ symmetry.
In the low density region
this chiral symmetry is broken down to its diagonal subgroup, but the
color symmetry is unbroken:
\begin{eqnarray}
{SU}(3)_{c} \times
{SU}(2)_{L}\times{SU}(2)_{R}
\rightarrow
{SU}(3)_{c} \times {SU}(2)_{V}
\ .
\nonumber
\end{eqnarray}
In the high density region, on the other hand,
the chiral symmetry is restored but the color symmetry is broken down to
its subgroup:~\cite{Al}
\begin{eqnarray}
{SU}(3)_{c} \times
{SU}(2)_{L}\times{SU}(2)_{R}
\rightarrow
{SU}(2)_{c} \times 
{SU}(2)_{L}\times{SU}(2)_{R}
\ .
\nonumber
\end{eqnarray}

This paper is organized as follows.
In \S\ref{Preliminary}, we summarize the quark propagator, the gluon
propagator and the running coupling that we use in the present analysis.
Several approximations of the quark propagator are made.
We also give formulas for calculating the diquark condensate 
and the chiral condensate. 
In \S\ref{Effective potential and Schwinger-Dyson equation},
 we present the effective potential for the quark propagator and 
then derive the Schwinger-Dyson equation as a stationary condition of 
the effective potential.
Section~\ref{Numerical analysis} is the main part of this paper,
 where we give the results of the numerical analysis of
 the Schwinger-Dyson equation.
Finally, we give a summary and discussion
 in \S\ref{Summary and Discussion}.
In the appendices we summarize several intricate expressions and 
useful formulas.

\section{Preliminaries}
\label{Preliminary}
In this section we briefly present the quark propagator, the gluon
propagator and the running coupling that we use in the numerical
analysis.
In \S\ref{Nambu-Gorkov fields and Quark propagator},
 we introduce the eight-component Majorana spinor
(Nambu-Gorkov field) and give the general form of the full quark
propagator as a matrix in the Nambu-Gorkov space.
We obtain constraints on the full propagator from the parity and
time reversal invariances.
We further make several assumptions to restrict the form of the propagator.
The gluon propagator with screening mass effects is presented in
 \S\ref{Gluon propagator and the running couplinig}.
We also give the explicit form of the running coupling that we
use in our numerical analysis. 
We give formulas to calculate the quark-antiquark condensate and the
diquark condensate in \S\ref{Condensates}.

\subsection{Nambu-Gorkov fields and quark propagator}
\label{Nambu-Gorkov fields and Quark propagator}
Since we are interested in the phase structure of QCD, including the 
color superconducting phase, it is convenient to use 
the eight-component Majorana spinor (Nambu-Gorkov field)
 instead of the four-component Dirac spinors. The Nambu-Gorkov field is 
expressed as
\begin{eqnarray}
 \Psi=\frac{1}{\sqrt{2}}\left(
			 \begin{array}{c}
			  \psi \\
			  \psi^C 
			 \end{array}\right) \ ,
 \quad \psi^C=C\bar\psi^T \ ,
\label{NambuGorkov}
\end{eqnarray}
where $C=i\gamma^2\gamma^0$.
(We suppress the color and flavor indices until later in the paper.)
Using the Nambu-Gorkov basis, the inverse of the free quark propagator 
is given by
\begin{eqnarray}
 i{S_F^{(0)}}^{-1}(p)=
  \left(
   \begin{array}{cc}
    (p_0+\mu)\gamma^0-\vec{\gamma}\cdot\vec{p} & 0 \\
    0 & (p_0-\mu)\gamma^0-\vec{\gamma}\cdot\vec{p}
   \end{array}
 \right) \ .
\end{eqnarray}
Similarly, the full quark propagator is written in matrix form
 in the Nambu-Gorkov space as
\begin{eqnarray}
 S_F(p)&=&
  \left(
   \begin{array}{cc}
    S_{F11}(p) & S_{F12}(p) \\
    S_{F21}(p) & S_{F22}(p)
   \end{array}
 \right) \ .
\end{eqnarray}
The four components of the above full quark propagator 
are not independent.
{}From the relation $\psi^C=C\bar\psi^T$ it can be shown that they satisfy
\begin{eqnarray}
\label{NGconstraint2}
 &&\hspace{2cm}S_{F11}(p)=C[S_{F22}(-p)]^TC^{-1} \ , \\
 &&S_{F21}(p)=C[S_{F21}(-p)]^TC^{-1} \ , \qquad
  S_{F12}(p)=C[S_{F12}(-p)]^TC^{-1} \ ,
\label{NGconstraint} 
\end{eqnarray}
where $^T$ indicates the transposed matrix.
Furthermore, we easily find the following relations among the
 components of the inverse propagator:\cite{Bailin}
\begin{eqnarray}
 \label{hermit1}
 i\{S_{F}^{-1}(p)\}_{11}+i\{S_{F}^{-1}(p)\}_{22}
  &=&\gamma^0[i\{S_{F}^{-1}(p)\}_{11}+i\{S_{F}^{-1}(p)\}_{22}]^\dag\gamma^0 \ ,
  \\
 \label{hermit2}
 i\{S_{F}^{-1}(p)\}_{21}&=&\gamma^0[i\{S_{F}^{-1}(p)\}_{12}]^\dag\gamma^0 \ .
\end{eqnarray}
Here $^\dag$ indicates the hermitian conjugate matrix.

The QCD Lagrangian at finite density is not invariant under 
charge conjugation ($\cal{C}$) due to the existence of the chemical 
potential $\mu$, while it is invariant under time reversal
($\cal{T}$) and parity ($\cal{P}$).
Time reversal of the full quark propagator is given by
\begin{eqnarray}
 S_F(p^0,\vec{p})\mathop{\longrightarrow}_{\cal T}
  \tilde{T}[S_F(p^0,-\vec{p})]^T\tilde{T} \ ,
  \quad\tilde{T}=\left(
		   \begin{array}{cc}
		    T & 0 \\
		    0 & T \\
		   \end{array}
		 \right) \ ,
  \ T=i\gamma^1\gamma^3\gamma^0 \ ,
\label{timet}
\end{eqnarray}
and the parity transformation is expressed as
\begin{eqnarray}
 S_F(p^0,\vec{p})\mathop{\longrightarrow}_{\cal P}
  \Gamma^0S_F(p^0,-\vec{p})\Gamma^0 \ , 
  \quad \Gamma^0=\left(
		   \begin{array}{cc}
		    \gamma^0 & 0 \\
		    0 & -\gamma^0 \\
		   \end{array}
		 \right) \ .
\label{parityt}
\end{eqnarray}
For $\mu=0$, charge conjugation is not broken explicitly.
The charge conjugation transformation is expressed as
\begin{eqnarray}
 S_F(p)\mathop{\longrightarrow}_{\cal C}\tilde{C}[S_F(-p)]^T\tilde{C}^{-1} \ ,
  \quad \tilde{C}=\left(
		    \begin{array}{cc}
		     C & 0 \\
		     0 & C \\
		    \end{array}
		  \right) \ .
\label{charget}
\end{eqnarray}

Let us obtain constrants to the spinor, color and flavor
structures of the propagator.\footnote{
 When the chiral symmetry is not broken, the following
 discussion is essentially the same as that in Ref.~\citen{Pisarski}.}
According to the spinor structure, the inverse full quark 
propagator is generally parametrized by eight bases
:
\begin{eqnarray}
 \label{basis}
 \{\enskip 1 ,\enskip \gamma_5 ,\enskip \gamma_0 ,\enskip \gamma_5\gamma_0 ,
  \enskip \vec{\gamma}\cdot\vec{p} ,\enskip \gamma_5\vec{\gamma}\cdot\vec{p} ,
  \enskip \gamma_0\vec{\gamma}\cdot\vec{p} ,\enskip 
  \gamma_5\gamma_0\vec{\gamma}\cdot\vec{p} \enskip\} \ .
\end{eqnarray}
Let $\{S_F^{-1}(p)\}_{IJ}$ $(I,J=1,2)$ denote the $I$-$J$ Nambu-Gorkov
 component of the inverse full propagator.\footnote{Note that in our notation
 $\{S_F^{-1}\}_{IJ}\neq\{{S_F}_{IJ}\}^{-1}$.}
By using the above bases, general forms of $\{S_F^{-1}(p)\}_{11}$ and 
$\{S_F^{-1}(p)\}_{12}$ can be expressed as
\begin{eqnarray}
\label{inprop1}
 i\{S_{F}^{-1}(p)\}_{11}&=&-B(p)+A(p)(p_0+\mu)\gamma_0
 +C(p)\vec{\gamma}\cdot\vec{p}+D(p)
 \gamma_0\vec{\gamma}\cdot\vec{p} 
 \nonumber\\ && 
 +\Delta_{(5)}^+(p)\gamma_5\Lambda_p^++\Delta_{(5)}^-(p)\gamma_5\Lambda_p^-
 +\Xi_{(5)}^+(p)\gamma_5\gamma_0\Lambda_p^+
 +\Xi_{(5)}^-(p)\gamma_5\gamma_0\Lambda_p^- \ , \nonumber\\ \\
\label{inprop2}
 i\{S_{F}^{-1}(p)\}_{12}&=&\Delta^+(p)\gamma_5\Lambda_p^+
  +\Delta^-(p)\gamma_5\Lambda_p^-
  +\Xi^+(p)\gamma_5\gamma_0\Lambda_p^+
  +\Xi^-(p)\gamma_5\gamma_0\Lambda_p^- \nonumber\\ && 
  +B_{(5)}(p)+A_{(5)}(p)\gamma_0
  +C_{(5)}(p)\vec{\gamma}\cdot\vec{p}
  +D_{(5)}(p)\gamma_0\vec{\gamma}\cdot\vec{p} \ ,
\end{eqnarray}
where $\Lambda_p^{-}$ and $\Lambda_p^{+}$ are the projection operators for 
the quark and antiquark in the massless limit:
\begin{eqnarray}
 \Lambda_p^{\mp}&=&\frac{1}{2}\biggl(1\mp\frac{\gamma^0\vec\gamma\cdot\vec{p}}
  {\bar{p}}\biggr) \ .
\end{eqnarray}
It should be noted
that the above functions $B$, $A$, $\cdots$, etc., do not carry
 spinor indices, while they are still matrices in the color and flavor spaces.
We assume here that $O(3)$ symmetry (spatial rotation) is not
broken, so that $B$, $A$, $\cdots$, etc., are functions of $p_0$
and $\bar{p} \equiv \vert \vec{p} \vert$.
To avoid notational complexity we write $B(p)$ in place of 
$B(p_0,\bar{p})$.
In the following discussion, we often use $B(-p)$, which represents
$B(-p_0,\bar{p})$.

Equations~(\ref{NGconstraint2}) and (\ref{hermit2}) relate
 $\{S_{F}^{-1}(p)\}_{22}$ and $\{S_{F}^{-1}(p)\}_{21}$ to 
$\{S_{F}^{-1}(p)\}_{11}$ and $\{S_{F}^{-1}(p)\}_{12}$, respectively.
By using the notation in Eqs.~(\ref{inprop1}) and (\ref{inprop2}),
 $\{S_{F}^{-1}(p)\}_{22}$ and $\{S_{F}^{-1}(p)\}_{21}$ can be expressed as
\begin{eqnarray}
\label{inprop3}
 i\{S_{F}^{-1}(p)\}_{22}&=&C[i\{S_{F}^{-1}(-p)\}_{11}]^TC^{-1} \nonumber\\
 &=&-[B(-p)]^T+[A(-p)]^T(p_0-\mu)\gamma_0
 +[C(-p)]^T\vec{\gamma}\cdot\vec{p}+[D(-p)]^T
 \gamma_0\vec{\gamma}\cdot\vec{p} 
 \nonumber\\ &&
 +[\Delta_{(5)}^+(-p)]^T\gamma_5\Lambda_p^+
 +[\Delta_{(5)}^-(-p)]^T\gamma_5\Lambda_p^- \nonumber\\
 &&+[\Xi_{(5)}^+(-p)]^T\gamma_5\gamma_0\Lambda_p^-
 +[\Xi_{(5)}^-(-p)]^T\gamma_5\gamma_0\Lambda_p^+ \ , \\
\label{inprop4}
 i\{S_{F}^{-1}(p)\}_{21}&=&\gamma_0[i\{S_{F}^{-1}(p)\}_{12}]^\dag\gamma_0
  \nonumber\\
 =&-&[\Delta^-(p)]^\dag\gamma_5\Lambda_p^+
  -[\Delta^+(p)]^\dag\gamma_5\Lambda_p^-
  +[\Xi^+(p)]^\dag\gamma_5\gamma_0\Lambda_p^+
  +[\Xi^-(p)]^\dag\gamma_5\gamma_0\Lambda_p^- \nonumber\\
  &+&[B_{(5)}(p)]^\dag+[A_{(5)}(p)]^\dag\gamma_0
  +[C_{(5)}(p)]^\dag\vec{\gamma}\cdot\vec{p}
  -[D_{(5)}(p)]^\dag\gamma_0\vec{\gamma}\cdot\vec{p} \ . \nonumber\\
\end{eqnarray}

In the above expressions, the full quark propagator
 includes sixteen functions, which are still 
matrices in the color and flavor spaces. Here we obtain the constraints 
on these functions from Eqs.~(\ref{NGconstraint}) and (\ref{hermit1}).
First, Eq.~(\ref{NGconstraint}) leads to 
\begin{eqnarray}
 \label{dNGconstraint:1}
  &&\Delta^\pm(p)=[\Delta^\pm(-p)]^T ,\quad \Xi^\pm(p)=[\Xi^\mp(-p)]^T 
  \ ,
\end{eqnarray}
and
\begin{eqnarray}
 &&B_{(5)}(p)=[B_{(5)}(-p)]^T ,\quad A_{(5)}(p)=-[A_{(5)}(-p)]^T \ ,\nonumber\\
 &&C_{(5)}(p)=[C_{(5)}(-p)]^T ,\quad D_{(5)}(p)=[D_{(5)}(-p)]^T \ .
\label{dNGconstraint:2}
\end{eqnarray}
Second, the relation (\ref{hermit1}) leads to
\begin{eqnarray}
 \label{Dhermite1}
  &&(p_0+\mu)A(p)+(p_0-\mu)[A(-p)]^T=[(p_0+\mu)A(p)+(p_0-\mu)[A(-p)]^T]^\dag 
  \ , \nonumber\\
  &&B(p)+[B(-p)]^T=[B(p)+[B(-p)]^T]^\dag \ ,
   \quad C(p)+[C(-p)]^T=[C(p)+[C(-p)]^T]^\dag \ ,
   \nonumber\\ 
  &&iD(p)+i[D(-p)]^T=[iD(p)+i[D(-p)]^T]^\dag \ , 
\end{eqnarray}
and
\begin{eqnarray}
 \label{Dhermite2}
  &&i\Delta_{(5)}^+(p)+i[\Delta_{(5)}^+(-p)]^T=
  [i\Delta_{(5)}^-(p)+i[\Delta_{(5)}^-(-p)]^T]^\dag \ , \nonumber\\
 &&\Xi_{(5)}^+(p)+[\Xi_{(5)}^-(-p)]^T
  =[\Xi_{(5)}^+(p)+[\Xi_{(5)}^-(-p)]^T]^\dag \ .
\end{eqnarray}

Let us further constrain the forms of the functions using
 time reversal (${\cal T}$) and parity (${\cal P}$).
In various analyses, 
\cite{Barr,Bailin,Al,Evans:1999ek,Schafer:1999na}
 it has been shown that the most favorable
condensate carries even parity, though it was pointed out in Ref.~\citen{Pi4}
that a parity violating condensate can be formed.
In the present analysis, however, we use the SDE with a one-gluon exchange
kernel, so that there exists ${U(1)_A}$ symmetry in the system.
This ${U(1)_A}$ symmetry enables us to choose the vacuum with even
parity. In such a case, the number of the functions is reduced to eight:
\begin{eqnarray}
 && B_{(5)}(p)=A_{(5)}(p)=C_{(5)}(p)=D_{(5)}(p)=0 \ , \\
 && \Delta_{(5)}^+(p)=\Delta_{(5)}^-(p)=\Xi_{(5)}^+(p)=\Xi_{(5)}^-(p)=0 \ .
\end{eqnarray}
If $\cal{T}$ is not broken spontaneously\footnote
{Note that the effective potential shown in Eq.~(\ref{efpm})
 is invariant under ${\cal T}$ transformation. Then, 
 $\mbox{Im} \Delta^- = \mbox{Im} \Delta^+ =0$ is actually
 a trivial solution to the imaginary part of
 the SDEs (\ref{SD12}) and (\ref{SD21}).}, 
we obtain the following constraints on the functions allowed 
by $\cal{P}$ invariance:
\begin{eqnarray}
 && [A(p)]^T=A(p) \ , \quad [B(p)]^T=B(p) \ , \quad [C(p)]^T=C(p) \ , \quad
  [D(p)]^T=-D(p) \ , \nonumber\\
 && [\Delta^\pm(p)]^*=\Delta^\pm(p) \ , \quad 
  [\Xi^\pm(p)]^*=\Xi^\pm(p) \ . \nonumber\\
\label{Timecon}
\end{eqnarray}

Now, let us consider the flavor and color structures of $A$, $B$, $C$,
$D$, $\Delta^\pm$ and $\Xi^\pm$. 
As we discussed in the Introduction,
we consider the strange quark mass to be large enough to ignore the
strange quark in the formation of the diquark condensate of $u$ and
$d$ quarks.
Then color superconductivity is realized in the 2SC phase, where
the color ${SU}(3)_{c}$ symmetry is broken down to its subgroup 
${SU}(2)_{c}$. In the chiral symmetry broken ($\chi{SB}$) phase,
on the other hand, the chiral ${SU}(2)_{L}\times{SU}(2)_{R}$ symmetry 
is broken down to its diagonal subgroup 
${SU}(2)_{V}$. Functions in the full quark propagator
 are written to reflect these symmetry breaking structures.
Here we consider the color and flavor structures in the mixed phase,
where the color symmetry breaking and the chiral symmetry breaking
discussed above occur simultaneously:
${SU}(3)_{c}\times{SU}(2)_{L}\times{SU}(2)_{R}
\rightarrow{SU}(2)_{c}\times{SU}(2)_{V}$.
The pure 2SC phase [${SU}(2)_{L}\times{SU}(2)_{R}$ is left
unbroken] or pure $\chi{SB}$ phase [${SU}(3)_{c}$ is left
unbroken] can be realized in a certain limit of certain functions. 
The pure 2SC phase is realized for $B=D=\Xi^\pm=0$, and the pure
$\chi{SB}$ phase for $\Delta^\pm=\Xi^\pm=0$.
Under this breaking pattern, the possible color and flavor structures of
the functions are given by
\begin{eqnarray}
 A(p) , \ B(p) , \ C(p) , \ D(p)
  &\sim&\delta^{ij}(\delta^{ab}-\delta^{a3}\delta^{b3}) \ , \quad 
  \delta^{ij}\delta^{a3}\delta^{b3} \ , \nonumber\\
\label{struct:fc}
 \Delta^\pm(p) , \ \Xi^\pm(p)
  &\sim&\epsilon_{ij}\epsilon^{ab3} \ , \quad \delta_{ij}\epsilon^{ab3} \ ,
\end{eqnarray} 
where $(a,b)$ and $(i,j)$ are the color and flavor indices,
respectively, and we fix the direction of the color symmetry breaking in the
3-direction, without loss of generality.
Under the transpose of color and flavor indices, 
the function $D$ must be symmetric, as dictated by the symmetry
breaking pattern [see Eq.~(\ref{struct:fc})], but antisymmetric, as
dictated by time reversal invariance [see Eq.~(\ref{Timecon})], 
and hence the function $D$ has to vanish:
\begin{eqnarray}
 D(p)=0 \ . \label{Tinv}
\end{eqnarray}

Next we make several approximations of the structure of the quark
propagator. At $T=\mu=0$, the SDE with the
ladder approximation in the Landau gauge leads to the fact
that the wave function renormalization of the quark 
yields 1($A(p)=C(p)=1$).
Actually, this is necessary for consistency with QCD 
in the ladder approximation (see,~e.g.,~Ref.~\citen{MT}). 
For $T > 0$ and/or $\mu >0$ it is also necessary to keep $A(p)=C(p)=1$ 
by adopting, e.g., nonlocal gauge fixing.\cite{Kugo:1992pr}
In the high density region, however,
the deviation of the wave function renormalization from 1 have
been shown to be small for any choice of the gauge parameter.\cite{Hong} 
In addition, the relations $A(p){\neq}1$ and $C(p){\neq}1$ do not imply 
chiral symmetry breaking. 
Therefore, in this paper we ignore
 the deviation from 1 and regard $A(p)=C(p)=1$ as an approximate
solution for general $\mu$, even though we keep the Landau gauge.
Furthermore, we assume that the chirality even channel in the majorana
masses is dominant and ignore the chirality odd parts:
\begin{eqnarray}
 \label{approx1}
 \Xi^+(p)=\Xi^-(p)=0 \ .
\end{eqnarray}
In fact, in the high density region, where the chiral condensate vanishes,
two massless quarks with the same helicity condense in the scalar (J=0)
 channel(see,~Ref.~\citen{Pisarski}).
Finally, since the color and flavor structure of the diquark condensate
is given by $\epsilon_{ij}\epsilon^{ab3}$, we ignore the terms
proportional to $\delta_{ij}\epsilon^{ab3}$ in the function
$\Delta^\pm$: We fix the color and flavor structures of $\Delta^\pm$ as
\begin{eqnarray}
 \Delta^\pm(p)\sim\epsilon_{ij}\epsilon^{ab3} \ . 
\end{eqnarray}

{}From the above discussion, 
the inverse full quark propagator can be expressed as
\begin{eqnarray}
\label{inprop}
 iS_{F}(p)^{-1}&=&
 \left(
    \begin{array}{cc}
     (p_0+\mu)\gamma^0-\vec{\gamma}\cdot\vec{p}-B(p) & \Delta(p) \\
     \tilde\Delta(p) & (p_0-\mu)\gamma^0-\vec{\gamma}\cdot\vec{p}-B(-p) 
    \end{array}
  \right) \ , \nonumber\\ 
\end{eqnarray}
where\footnote{Note that in Eqs.~(\ref{inprop2}),
 (\ref{inprop4}), (\ref{dNGconstraint:1}), (\ref{Timecon}) and
 (\ref{struct:fc}),
 $\Delta^{+}$ and $\Delta^{-}$ are still matrices 
in the color and flavor spaces.
In Eq.~(\ref{massf}) the same notation is
used to express the scalar functions
 after factoring the color and flavor indices.}
\begin{eqnarray}
 &&\hspace{1cm}\Delta(p)^{ab}_{ij}=\epsilon_{ij}\epsilon^{ab3}\gamma_5
  [\Delta^{+}(p)\Lambda_p^{+}+\Delta^{-}(p)\Lambda_p^{-}] \ ,\nonumber\\
 &&\hspace{1cm}\tilde\Delta(p)^{ab}_{ij}
  =\gamma^0{\Delta(p)^{\dag}}^{ab}_{ij}\gamma^0 \nonumber\\
 &&\hspace{2.1cm}=-\epsilon_{ij}\epsilon^{ab3}\gamma_5
  [\Delta^{+}(p)\Lambda_p^{-}+\Delta^{-}(p)\Lambda_p^{+}] \ ,\nonumber\\
 &&\hspace{1cm}B(p)^{ab}_{ij}=B_1(p)\delta^{ij}
  (\delta^{ab}-\delta^{a3}\delta^{b3})
  +B_3(p)\delta^{ij}\delta^{a3}\delta^{b3} \ .\label{massf}
\end{eqnarray}
Now the full quark propagator includes four scalar functions,
$B_{1}$ and $B_{3}$ corresponding to the Dirac masses 
responsible for the chiral symmetry breaking
and $\Delta^{+}$ and $\Delta^{-}$ corresponding to the 
Majorana masses responsible for the color symmetry breaking.
{}From Eq.~(\ref{Dhermite1}), the Dirac masses $B_{1}$ and $B_{3}$ obey
the constraint
\begin{eqnarray}
 \label{diraprop}
  \mbox{Im} \left[ B_{i}(p) \right] = - \mbox{Im}
  \left[ B_{i}(-p) \right] \ . \qquad\ ( i = 1,3 ) 
\end{eqnarray}
Equations (\ref{dNGconstraint:1}) and (\ref{Timecon}) imply
 that the Majorana masses $\Delta^{+}(p)$ and $\Delta^{-}(p)$ are real 
and even functions of $p_0$:
\begin{eqnarray}
 \label{majprop}
  \Delta^{\pm}(p) = \Delta^{\pm}(-p) = 
  \left[ \Delta^{\pm}(p) \right]^{\ast} \ .
\end{eqnarray}

By taking the inverse of the expression in Eq.~(\ref{inprop}),
the full quark propagator is given by \cite{Hong}
\begin{eqnarray}
 \label{quarkp}
 -iS_{F}(p)&=&\left(
  \begin{array}{l}
   \hspace{0.5cm} R^{-1}_+(p) \hspace{1.7cm} 
    -\{(p_0+\mu)\gamma^0-\vec{\gamma}\cdot\vec{p}-B(p)\}^{-1}
    \Delta(p)R^{-1}_-(p) \\
   -\{(p_0-\mu)\gamma^0-\vec{\gamma}\cdot\vec{p}-B(-p)\}^{-1}
  \tilde\Delta(p)R^{-1}_+(p) \hspace{1.6cm} R^{-1}_-(p) 
  \end{array}
  \right) \ , \nonumber\\
\end{eqnarray}
where
\begin{eqnarray}
 R_+(p)&\equiv&\{(p_0+\mu)\gamma^0-\vec{\gamma}\cdot\vec{p}-B(p)\}
  -\Delta(p)\{(p_0-\mu)\gamma^0-\vec{\gamma}\cdot\vec{p}-B(-p)\}^{-1}
  \tilde\Delta(p) \ ,\nonumber\\
 R_-(p)&\equiv&\{(p_0-\mu)\gamma^0-\vec{\gamma}\cdot\vec{p}-B(-p)\}
  -\tilde\Delta(p)\{(p_0+\mu)\gamma^0-\vec{\gamma}\cdot\vec{p}-B(p)\}^{-1}
  \Delta(p) \ .\nonumber\\
\end{eqnarray}
To use $S_F$ in the Schwinger-Dyson equation,
we need explicit forms of $R_+^{-1}$, $R_-^{-1}$, and so on.
We list them in
 Appendix~\ref{Quark Propagator and the Schwinger-Dyson Equation}.

\subsection{Gluon propagator and the running coupling}
\label{Gluon propagator and the running couplinig}
In dense medium, gluons generally acquire Debye masses.
On the basis of the renormalization group equation\cite{Son}
 and the Schwinger-Dyson equation,\cite{Hong,Schaf3} in the high
density region it has been shown that the long range interaction 
mediated by magnetic gluons gives 
 the dominant contribution to the formation of the Majorana mass gap. 
The value of the Majorana mass gap obtained from these analyses 
is consistent with that derived using models based on the instanton
induced interaction.\cite{Al}
In addition to the Debye masses,
five gluons have Meissner masses, due to the Meissner-Higgs effect,
when the color $SU(3)$ is broken to its subgroup $SU(2)$.
According to the result of the SDE analysis in the high density
 region,\cite{Hong} these masses are of no importance in the SDE.

In this paper we include the Debye screening and the Landau damping
 in the gluon propagator through the hard dense loop approximation.
However, we ignore Meissner masses.
Moreover, as explained in the previous subsection,
we take the Landau gauge for the gluon propagator
and assume $A(p)=C(p)=1$ which is needed for consistency 
with QCD in the ladder approximation 
(see, e.g., Ref.~\citen{Kugo:1992pr}).

The explicit form of the gluon propagator that we use in this paper
is given by~\cite{Hong}
\begin{eqnarray}
 D_{\mu\nu}^{AB}(k)&\equiv&{\delta^{AB}D_{\mu\nu}(k)} \nonumber\\
  &=&i\delta^{AB}\frac{\vert\vec{k}\vert}
  {\vert\vec{k}\vert^3+\pi{M_D}^2\vert{k_4}\vert/2}
  O^{(1)}_{\mu\nu}
  +i\delta^{AB}\frac{1}{(k_4)^2+\vert\vec{k}\vert^2+2M_D^2}
  O^{(2)}_{\mu\nu} \ , \nonumber\\
\label{gluonp}
\end{eqnarray}
where $k_4=-ik_0$ and $M_D$ is the Debye mass of the gluon.
In the hard dense loop approximation this is given by~\cite{Hong} 
\begin{eqnarray}
 M_D=\frac{\sqrt{N_f}}{2\pi}g(\mu)\mu \ .
\end{eqnarray}
$O^{(i)}_{\mu\nu}(i=1,2)$ are the polarization tensors defined by
\begin{eqnarray}
 &&O^{(1)}_{\mu\nu}=P^\bot_{\mu\nu}+\frac{(u\cdot{k})^2}{(u\cdot{k})^2-k^2}
  P^u_{\mu\nu} \ , \quad 
  O^{(2)}_{\mu\nu}=-\frac{(u\cdot{k})^2}{(u\cdot{k})^2-k^2}P^u_{\mu\nu} \ ,
\end{eqnarray}
where
\begin{eqnarray}
 &&P^\bot_{\mu\nu}=g_{\mu\nu}-\frac{k_\mu{k}_\nu}{k^2} \ ,
  \quad 
  P^u_{\mu\nu}=\frac{k_\mu{k}_\nu}{k^2}-\frac{k_\mu{u}_\nu+u_\mu{k}_\nu}
  {u\cdot{k}}+\frac{u_\mu{u}_\nu}{(u\cdot{k})^2}k^2 \ .
\end{eqnarray}
The Lorentz four-vector $u^\mu = (1,0,0,0)$ 
in the gluon propagator reflects the explicit breaking of Lorentz
symmetry due to the existence of the chemical potential 
in the rest frame of the medium.

In the SDE at zero density, it is important to use the running coupling, 
since the high-energy behavior of the mass function derived from 
the SDE with the running coupling is consistent with that derived from 
the operator product expansion.\cite{MT}
One consistent way to include the effect of the running
coupling is to use the improved ladder 
approximation,\cite{HM}
in which the high-energy behavior of the running coupling is
determined by the one-loop renormalization group equation derived in
QCD and the low-energy behavior is suitably regularized.
In the present analysis we consider low and medium 
density regions, so that we use the
Higashijima-Miransky-type\cite{HM,Ao} 
running coupling
\begin{eqnarray}
\label{run}
 &&\alpha_s(E)=\frac{g^2(E)}{4\pi}=
  \frac{6\pi/(11N_c-2N_f)}{\max(t,t_f)} \ , 
\end{eqnarray}
where
\begin{eqnarray}
\label{scale}
 && t=\ln\frac{E}{\Lambda_{\rm qcd}} \ , \qquad 
  t_f=\ln\frac{E_f}{\Lambda_{\rm qcd}} \ , \qquad 
\end{eqnarray}
with $E$ being the energy scale, $\Lambda_{\rm qcd}$
 the characteristic scale of QCD,\footnote{Here, $\Lambda_{\rm qcd}$ is 
determined from the infrared structure in the present analysis, while 
the usual $\Lambda_{\rm QCD}$ is determined from the ultraviolet structure.}
 and $E_f$ the infrared cutoff scale introduced
 to regularize the infrared singularity.\footnote{
 When we study the high density region, we need to add another cutoff scale,
 $\mbox{ln}(\mu/\Lambda_{\rm qcd})$, at which the running stops.
 However, this is neglected here, because we study only the intermediate
 region around the phase transition point.}
 As discussed in the Introduction,
 in the present analysis we assume that the mass of the strange quark is 
large enough to ignore the $s$-quark in the formation of the diquark 
condensation of $u$ and $d$ quarks.
On the other hand, it is natural to assume that the current mass of 
the $s$-quark is
 smaller than $\Lambda_{\rm qcd}$. In such a case,
the effect from the $s$-quark should be included 
in the running coupling.\footnote{As we show in the next section, 
$\Lambda_{\rm qcd}={604}$\,MeV in the present analysis,
 which is obviously larger than the $s$-quark mass.}
Thus we set $N_f=3$ and $N_c=3$ in the running coupling (\ref{run}). 
Note that the value of the running coupling in the infrared region 
needs to be sufficiently large in order to involve the effect of dynamical 
symmetry breaking, i.e., $t_f$ has to be sufficiently small.
Here we set $t_f=0.5$, around which the various physical quantities for $\mu=0$
are very stable with respect to the changes in $t_f$.\cite{Ao}
In the numerical analysis we investigated the $t_f$ dependence of the results. 

\subsection{Condensates}
\label{Condensates}
In this subsection we give formulas to calculate
 the chiral condensate
 and the diquark condensate. 

The chiral condensate is generally expressed as
\begin{eqnarray}
 &&\langle{\Omega}\vert\bar\psi_{a}^i\psi^{a}_i(0)\vert{\Omega}
  \rangle_{\Lambda}=-\int^\Lambda\frac{d^4p}{(2\pi)^4}
  \mbox{tr}[S_{F11}] \ , 
\label{Conbqq}
\end{eqnarray}
 where $\vert{\Omega}\rangle$ is the ground state at nonzero density
 and the trace is taken in the spinor, flavor and color spaces.
Summations over the color index $a$ and the flavor index $i$ are
implicitly taken on the left-hand side of Eq.~(\ref{Conbqq}).
The quantity $\Lambda$ is the ultraviolet cutoff, introduced to regularize
 the logarithmic divergence.
In the actual numerical analysis we introduced two cutoffs 
for the temporal and spatial components of the momentum,
 but they are expressed symbolically as $\Lambda$ in this subsection.
For the present form of the full quark propagator in Eq.~(\ref{quarkp}),
this is given by
\begin{eqnarray}
 &&\langle{\Omega}\vert\bar\psi_{a}^i\psi_{i}^a(0)
  \vert{\Omega}\rangle_{\Lambda} \nonumber\\
 &&\hspace{1cm}=
  \int^\Lambda\frac{d^4p}{i(2\pi)^4}\nonumber\\
 &&\hspace{1.5cm}\biggl[\frac{16}{F(p,B_1,\Delta)}
  \biggl\{\biggl((p_0-\mu)^2-\bar{p}^2-\{B_1(-p)\}^2\biggr)B_1(p)
  -\Delta^+(p)\Delta^-(p)B_1(-p)\biggr\} \nonumber\\
  &&\hspace{2cm}+\frac{8}{F(p,B_3,\Delta=0)}
  \biggl((p_0-\mu)^2-\bar{p}^2-\{B_3(-p)\}^2\biggr)B_3(p)
  \biggr] \ , 
\end{eqnarray}
where $F$ is defined as
\begin{eqnarray}
  F(p,B_1,\Delta)&=&[(p_0+\mu)^2-\bar{p}^2-\{B_1(p)\}^2]
  [(p_0-\mu)^2-\bar{p}^2-\{B_1(-p)\}^2] \nonumber\\
 &&-[(p_0)^2-(\bar{p}-\mu)^2]\vert\Delta^+(p)\vert^2
  -[(p_0)^2-(\bar{p}+\mu)^2]\vert\Delta^-(p)\vert^2 \nonumber\\
 &&+\vert\Delta^+(p)\vert^2\vert\Delta^-(p)\vert^2
  +2B_1(p)B_1(-p)\Delta^+(p)\Delta^-(p) \ .
\label{Fdef}
\end{eqnarray}
For the chiral condensate of the charge conjugated quarks,
 we obtain the following relation from Eq.~(\ref{NGconstraint2}):
\begin{eqnarray}
 &&\langle{\Omega}\vert[\bar\psi_C]_{a}^i[\psi_C]_{i}^a(0)
  \vert{\Omega}\rangle_{\Lambda} \nonumber\\
 &&\hspace{1cm}=-\int^\Lambda\frac{d^4p}{(2\pi)^4}
  \mbox{tr}[S_{F22}]
  =-\int^\Lambda\frac{d^4p}{(2\pi)^4}
  \mbox{tr}[S_{F11}] \nonumber\\
 &&\hspace{1cm}=
  \langle{\Omega}\vert\bar\psi_{a}^i\psi_{i}^a(0)
  \vert{\Omega}\rangle_{\Lambda} \ .
\end{eqnarray}
These expressions take the following familiar form
 when we set $\mu=0$, $\Delta=0$ and $B_1=B_3=B$:
\begin{eqnarray}
 \langle{\Omega}\vert\bar\psi_{a}^i\psi_{i}^a(0)
  \vert{\Omega}\rangle_{\Lambda}
  &=&\langle{\Omega}\vert[\bar\psi_C]_{a}^i[\psi_C]_{i}^a(0)
  \vert{\Omega}\rangle_{\Lambda} \nonumber\\
 &=&-4N_cN_f\int^\Lambda\frac{d^4p}{i(2\pi)^4}
  \frac{B(p)}{-(p_0)^2+\bar{p}^2+\{B(p)\}^2} \ ,
\end{eqnarray}
where $N_c=3$ and $N_f=2$.

The diquark condensate is generally expressed as
\begin{eqnarray}
 &&\langle{\Omega}\vert(\epsilon^{ij}\epsilon_{ab3})
  [\psi^T]_i^aC\gamma_5\psi_j^b(0)\vert{\Omega}\rangle_{\Lambda}
  =-\int^\Lambda\frac{d^4p}{(2\pi)^4}  
  \mbox{tr}[\epsilon^{(c)}\epsilon^{(f)}S_{F12}\gamma_5] \ , 
\end{eqnarray}
where $\epsilon^{(c)}$ and $\epsilon^{(f)}$ are antisymmetric matrices 
in the color and flavor spaces, respectively:
\begin{eqnarray}
 \{\epsilon^{(c)}\}^{ab}=\epsilon^{ab3} \ , \hspace{1.5cm}
  \{\epsilon^{(f)}\}_{ij}=\epsilon_{ij} \ . 
\end{eqnarray}
In the present approximation, the diquark condensate is given by
\begin{eqnarray}
 &&\langle{\Omega}\vert(\epsilon^{ij}\epsilon_{ab3})
  [\psi^{T}]^a_iC\gamma_5\psi_j^b(0)\vert{\Omega}\rangle_{\Lambda} \nonumber\\
 &&\hspace{1cm}=\int^\Lambda\frac{d^4p}{i(2\pi)^4}\frac{8}{F(p,B_1,\Delta)}
  \biggl[\biggl\{(p_0)^2-(\bar{p}-\mu)^2-\{\Delta^-(p)\}^2
  -\vert{B_1}(p)\vert^2\biggr\}\Delta^+(p) \nonumber\\
 &&\hspace{5cm}+\biggl\{(p_0)^2-(\bar{p}+\mu)^2-\{\Delta^+(p)\}^2
  -\vert{B_1}(p)\vert^2\biggr\}\Delta^-(p)\biggr] \ . \nonumber\\
\end{eqnarray}
Using time reversal invariance [see Eq.~(\ref{timet})], 
we obtain 
\begin{eqnarray}
 &&\langle{\Omega}\vert(\epsilon^{ij}\epsilon_{ab3})
  [\psi_C^T]_i^aC\gamma_5[\psi_C]_j^b(0)\vert{\Omega}\rangle_{\Lambda}
  \nonumber\\
 &&\hspace{1cm}=
  -\int^\Lambda\frac{d^4p}{(2\pi)^4}  
  \mbox{tr}[\epsilon^{(c)}\epsilon^{(f)}S_{F21}\gamma_5]
  =\int^\Lambda\frac{d^4p}{(2\pi)^4}
  \mbox{tr}[\epsilon^{(c)}\epsilon^{(f)}S_{F12}\gamma_5] \nonumber\\
 &&\hspace{1cm}=-\langle{\Omega}\vert
  [\psi^T]_i^aC\gamma_5\psi_j^b(0)\vert{\Omega}\rangle_{\Lambda} \ . 
\end{eqnarray}
We note that the parity invariance existing in the present analysis
 leads to a vanishing parity violating condensate: 
\begin{eqnarray}
 &&\langle{\Omega}\vert(\epsilon^{ij}\epsilon_{ab3})
  [\psi^T]_i^aC\psi_j^b(0)\vert{\Omega}\rangle_{\Lambda}
  =-\int^\Lambda\frac{d^4p}{(2\pi)^4}  
  \mbox{tr}[\epsilon^{(c)}\epsilon^{(f)}S_{F12}]=0 \ . 
\end{eqnarray}

In the improved ladder approximation at zero density,
the high-energy behavior of the mass function is consistent with that
derived using the operator product expansion (OPE).
The chiral condensate calculated using the mass function 
was shown to obey the renormalization group evolution derived
with the OPE (see, e.g., Refs.~\citen{MT}).
Then, we identify the condensates, which are calculated with cutoff $\Lambda$, 
with those renormalized at the scale $\Lambda$ in QCD.
Therefore we scale them to the condensates at 1 GeV, using the leading 
renormalization group formulas. 
The relation between the chiral condensate at the scale $\Lambda$ and
that at the scale $E$ is given by
\begin{eqnarray}
 \label{bqqrenom}
 &&\langle{\Omega}\vert\bar\psi_{a}^i\psi^{a}_i(0)\vert{\Omega}
  \rangle_E
  =\biggl[\frac{\ln{E/\Lambda_{\rm qcd}}}
  {\ln{\Lambda/\Lambda_{\rm qcd}}}\biggr]^\kappa
  \langle{\Omega}\vert\bar\psi_{a}^i\psi^{a}_i(0)\vert{\Omega}\rangle_{\Lambda}
  \ , 
\end{eqnarray}
where
\begin{eqnarray}
 &&\kappa=\frac{9C_2(F)}{11N_c-2N_f}=
  \frac{9}{11N_c-2N_f}\frac{N_c^2-1}{2N_c} \ . 
  \label{AD}
\end{eqnarray}
Noting that
the attractive force between two quarks in the $\bar{\bf3}$ channel
due to one-gluon exchange is one half of that between a quark and
antiquark in the singlet channel,
we immediately obtain the following relation between the diquark condensate
 at the scale $\Lambda$ and that at the scale $E$:
\begin{eqnarray}
 \label{qqrenom}
 &&\langle{\Omega}\vert(\epsilon^{ij}\epsilon_{ab3}) 
  [\psi^T]_i^aC\psi_j^b(0)\vert{\Omega}\rangle_E
  =\biggl[\frac{\ln{E/\Lambda_{\rm qcd}}}
  {\ln{\Lambda/\Lambda_{\rm qcd}}}\biggr]^{\kappa/2}
  \langle{\Omega}\vert(\epsilon^{ij}\epsilon_{ab3}) 
  [\psi^T]_i^aC\psi_j^b(0)\vert{\Omega}\rangle_\Lambda
  \ . \nonumber\\
\end{eqnarray}

\section{Effective potential and the Schwinger-Dyson equation}
\label{Effective potential and Schwinger-Dyson equation}
In this section
we present the effective potential for the quark propagator and then
derive the Schwinger-Dyson equation (SDE) as a stationary condition of
the effective potential.

As explained in the Introduction,
we assume that the current mass of the strange quark is large 
enough to ignore it in the valence quark sector, and
we include only $u$ and $d$ quarks in the effective action.
Then, the effective action for the full quark propagator $S_F$ is
given by 
\cite{Corn}
\begin{eqnarray}
\label{Efa}
 \Gamma[S_F]&=&\frac{1}{2}\biggl(-i\mbox{Tr}\mbox{Ln}(S_F^{-1})
 -i\mbox{Tr}({S^{(0)}_F}^{-1}S_F)
  -i\Gamma_{\rm{2PI}}\it[S_F]\biggr) \ , 
\end{eqnarray}
where $\mbox{Tr}$ and $\mbox{Ln}$ are taken in all the spaces and  
$\Gamma_{\rm{2PI}}\it[S_F]$ represents the contributions 
from the two-particle irreducible (with respect to the quark line) diagrams.
The factor of 1/2 appears because we are using the eight-component 
Nambu-Gorkov spinor basis. 
In the high density region, the one-gluon exchange approximation is valid, 
since the coupling is weak.
In the present analysis, we extrapolate this approximation to  
intermediate density and include only the contribution from
 the one-gluon exchange diagram in $\Gamma_{\rm 2PI}[S_F]$,
\begin{eqnarray}
 \Gamma_{\rm 2PI}[S_F]=-\frac{1}{2}
  \mbox{Tr}(S_F\cdot{ig}\Gamma^\mu_A\cdot{S_F}\cdot{ig}\Gamma^\nu_B
  \cdot{D}_{\mu\nu}^{AB}) \ ,
\end{eqnarray} 
where $\Gamma^\mu_A$ is the quark-gluon vertex in the Nambu-Gorkov basis,
defined as
\begin{eqnarray}
 \Gamma^\mu_A=
  \left(
   \begin{array}{cc} 
    \gamma^\mu{T}_A & 0 \\
    0 & -\gamma^\mu(T_A)^T
   \end{array}
 \right) \ .
\end{eqnarray}
{}From the effective action (\ref{Efa}), the effective potential in 
the momentum space can be written 
\begin{eqnarray}
 \label{efpm}
 V[S_F]&=&-\Gamma[S_F]/\int{d^4x} \nonumber\\
 &=&\frac{1}{2}\int\frac{d^4p}{i(2\pi)^4}
  \biggl(\mbox{ln det}\{S_F(p)\}-\mbox{tr}\{S_F^{(0)-1}(p)S_F(p)\}\biggr)
  \nonumber\\&&
  +\frac{1}{2}\int\frac{d^4p}{i(2\pi)^4}\int\frac{d^4q}{i(2\pi)^4}
  \frac{1}{2}\mbox{tr}\{S_F(p)\cdot{ig\Gamma_A^\mu}\cdot{S_F(q)}\cdot
  {ig\Gamma_B^\nu}\}\cdot{iD^{AB}_{\mu\nu}(p-q)} \ , \nonumber\\
\end{eqnarray}
where ln, det and tr are taken in the spinor, color and flavor spaces.

The SDE is obtained as 
the stationary condition of the effective potential \\
($\delta{V}[S_F]/\delta{S_F}=0$):
\begin{eqnarray}
 \label{SD}
 S_F^{-1}&=&{S^{(0)}_F}^{-1}-
  (ig\Gamma^\mu_A\cdot{S_F}\cdot{ig}\Gamma^\nu_B)\cdot{D}_{\mu\nu}^{AB} \ .
\end{eqnarray}
In principle,
Eq.~(\ref{SD}) leads to coupled equations for seven
functions [$B$, $A$, $C$, $\Delta^{\pm}$ and $\Xi^{\pm}$ in 
Eqs.~(\ref{inprop1})--(\ref{inprop4})] in the quark propagator, 
which are still matrices in the color and flavor spaces.
As discussed in \S\ref{Nambu-Gorkov fields and Quark
propagator}, we assume that $A=C=1$ and $\Xi^{\pm}=0$ are
 the approximate solutions of the equations.
Furthermore, in the present pattern of the symmetry breaking, we take 
the color and flavor structures of the Dirac and Majonara masses 
as to be those given in Eq.~(\ref{massf}).
As a result, the above SDE~(\ref{SD}) leads to four coupled equations
for four scalar functions, $B_{1}$, $B_{3}$, $\Delta^{+}$ 
and $\Delta^{-}$. These are given by
\begin{eqnarray}
 \label{SD11a}
  B_1(p)&=&\int\frac{d^4q}{i(2\pi)^4}\frac{1}{2}
  \pi\alpha_s{D}_{\mu\nu}(q-p) 
  \mbox{tr}[\gamma^{\mu}T^AS_{F11}(q)\gamma^{\nu}T^A\delta^{(c)}_1] \ , \\
 \label{SD11b}
  B_3(p)&=&\int\frac{d^4q}{i(2\pi)^4}\pi\alpha_s{D}_{\mu\nu}(q-p) 
  \mbox{tr}[\gamma^{\mu}T^AS_{F11}(q)\gamma^{\nu}T^A\delta^{(c)}_3] \ , 
\end{eqnarray}
\newpage
\begin{eqnarray}
 \label{SD12} 
  \Delta^-(p)&=&\int\frac{d^4q}{i(2\pi)^4}\frac{1}{2}
  \pi\alpha_s{D}_{\mu\nu}(q-p) 
  \mbox{tr}[\gamma^{\mu}T^AS_{F12}(q)\gamma^{\nu}(T^A)^T\Lambda_p^-\gamma_5
  \epsilon^{(c)}\epsilon^{(f)}] \ , \nonumber \\ && \\
 \label{SD21} 
  \Delta^+(p)&=&\int\frac{d^4q}{i(2\pi)^4}\frac{1}{2}
  \pi\alpha_s{D}_{\mu\nu}(q-p) 
  \mbox{tr}[\gamma^{\mu}T^AS_{F12}(q)\gamma^{\nu}(T^A)^T\Lambda_p^+\gamma_5
  \epsilon^{(c)}\epsilon^{(f)}] \ , \nonumber \\ &&
\end{eqnarray}
where
\begin{eqnarray}
 \{\delta^{(c)}_1\}^{ab}=\delta^{ab}-\delta^{a3}\delta^{b3} \ , \hspace{1.5cm}
  \{\delta^{(c)}_3\}^{ab}=\delta^{a3}\delta^{b3} \ ,  
\end{eqnarray}
and the traces are taken in the spinor, flavor and color spaces.
The quantity $\alpha_s=\alpha_s(E)$ on the right-hand sides of
Eqs.~(\ref{SD11a})--(\ref{SD21}) is the running coupling defined in
 Eq.~(\ref{run}). For consistency with the chiral symmetry,
the argument of the running coupling should be chosen as
$E^2=-(p-q)^2$.\cite{Kugo:1992pr} 
However, it has been shown~\cite{Kugo:1992pr}
that the angular averaged form $E^2=-p^2-q^2$ is a
good approximation at $T=\mu=0$. 
For this reason, in the present analysis we use 
the angular averaged form for $\mu>0$.
Graphical representations of these gap equations are displayed in Fig.~1, 
and the explicit forms of these SDEs are given in
 Appendix~\ref{Quark Propagator and the Schwinger-Dyson Equation}.

\vspace{0.5cm}

\begin{minipage}{6cm}
 \epsfxsize=6cm
  \epsfbox{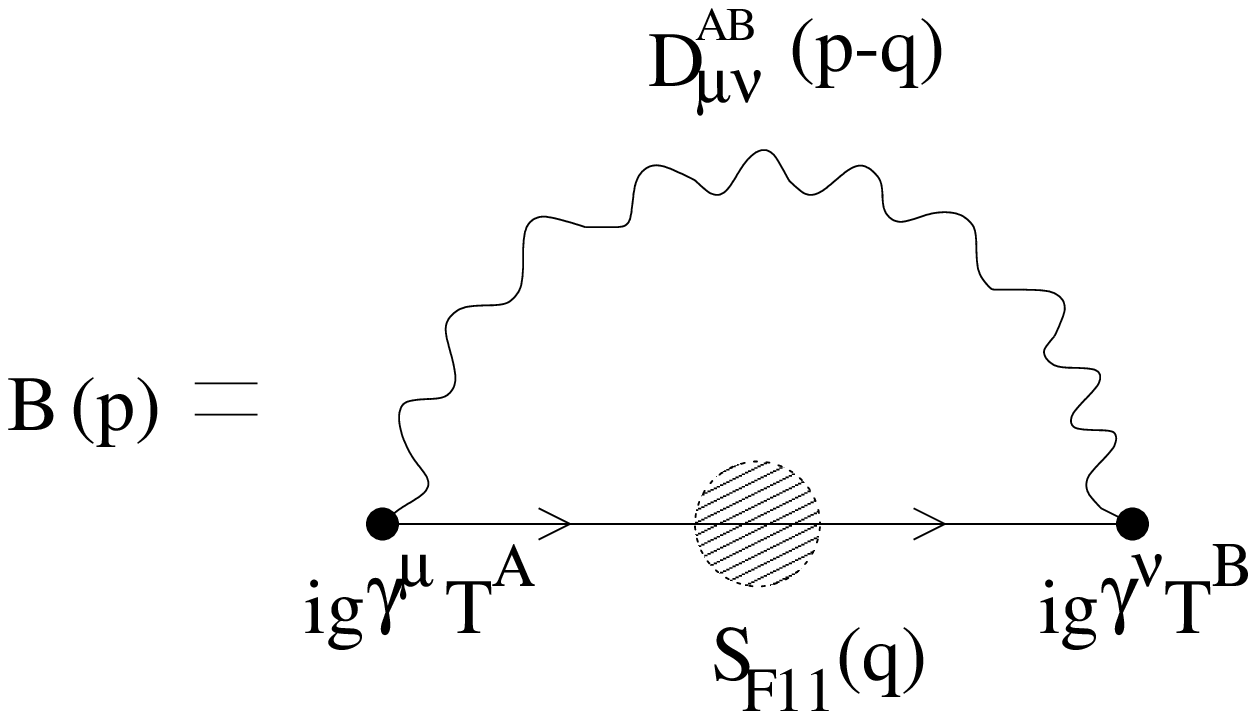} 
\end{minipage}
\hspace{0.5cm}
\begin{minipage}{6cm}
 \epsfxsize=6cm
  \epsfbox{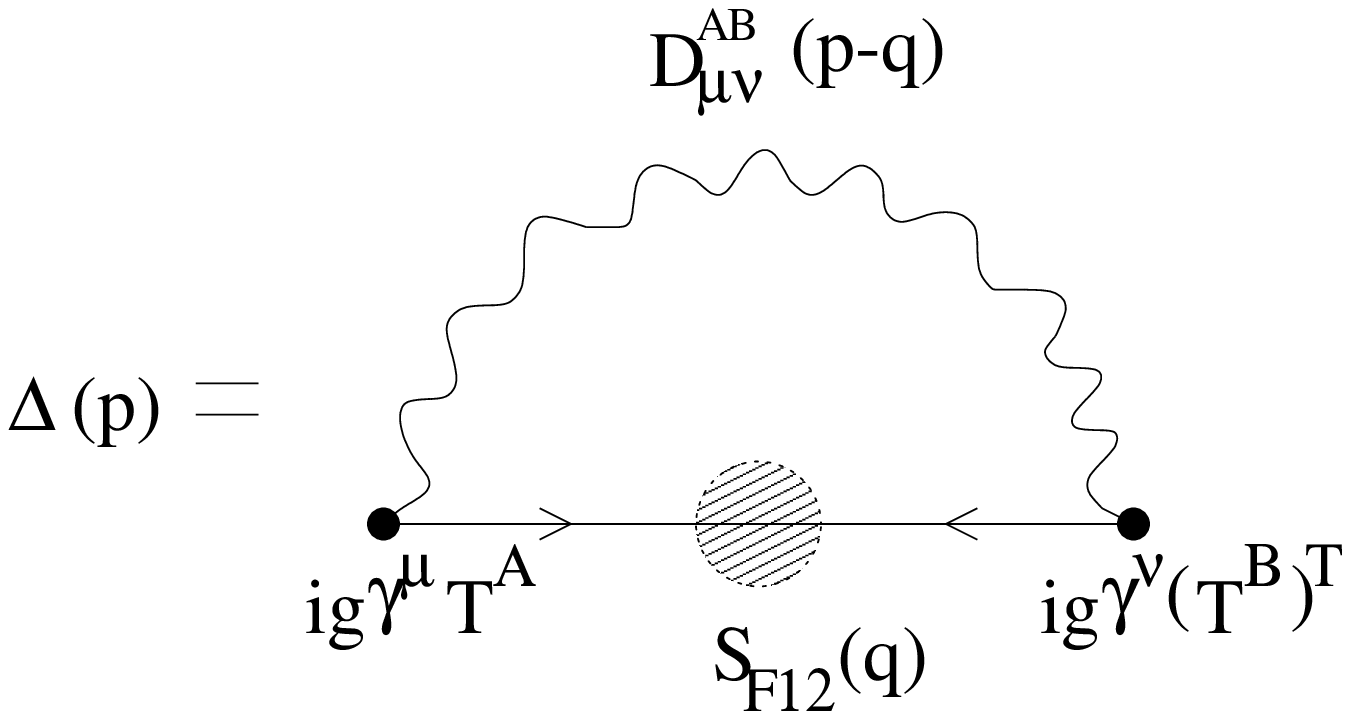}
\end{minipage} 

\vspace{0.3cm}

Fig.~1$\cdot$ Graphical representations of the gap equations.

\vspace{0.5cm}

As we show in Eq.~(\ref{majprop}), the Majonara masses
$\Delta^{+}(p)$ and $\Delta^{-}(p)$ are real and even functions of $p_0$,
while in general the Dirac masses $B_{1}(p)$ and $B_{3}(p)$ are 
complex functions.
Equation~(\ref{diraprop}) is the constraint on the imaginary part, while
no constraint is obtained on the real part from general
considerations.
However, as shown in Refs.~\citen{Tani} and \citen{Ha} for the case of
$\Delta^{+}(p)=\Delta^{-}(p)=0$, the structure of the SDE leads to a natural
constraint on the real parts of the Dirac masses $B_{1}(p)$ and $B_{3}(p)$.
This is shown as follows.
Let us take the complex conjugate of the SDEs
for $B_{1}$ and $B_{3}$ in Eqs.~(\ref{SDm}) and (\ref{SDm3}).
By using the fact that the kernel $K_0$ and the Majorana masses
$\Delta^{+}$ and $\Delta^{-}$ are real functions,
it is easily shown that the combination 
$\{ B_1^\ast(p) , B_3^\ast(p) \}$ satisfies the same coupled equations
as $\{ B_1(p^\ast) , B_3(p^\ast) \}$.
Since $p^{\mu\ast} = (p^{0\ast},\vec{p}) = 
(-ip_4,\vec{p}) = (-p^0,\vec{p})$, this means that
$\{ B_1^\ast(p) , B_3^\ast(p) \}$ satisfies the same coupled equations
as $\{ B_1(-p) , B_3(-p) \}$.
Thus these two combinations are equal up to their signs,
$\{ B_1^\ast(p) , B_3^\ast(p) \} =
\{ B_1(-p) , B_3(-p) \}$
or
$\{ B_1^\ast(p) , B_3^\ast(p) \} = -
\{ B_1(-p) , B_3(-p) \}$.
Since the general relation (\ref{diraprop}) implies that 
imaginary parts of $B_1(p)$ and $B_3(p)$ are odd functions of $p^0$,
the above argument implies that
the real parts are even functions of $p^0$:
\begin{eqnarray}
\label{diratprop}
B_{1,3}(-p) = B_{1,3}^\ast(p) \ .
\end{eqnarray}
Using the above properties for $B_{1}$ and $B_{3}$ and those 
for $\Delta^{+}$ and $\Delta^{-}$ in Eq.~(\ref{majprop}), 
we can always restrict the integration over the
temporal component of the momentum, $p_4 = -i p^0$, to its positive
region.

Substituting the solution of Eq.~(\ref{SD}) into Eq.~(\ref{efpm}),
 we obtain the effective potential 
at the vacuum, i.e.,~at the stationary point. 
Since the effective potential itself is divergent,
 we subtract the effective potential at the trivial vacuum:
\begin{eqnarray}
 &&\bar{V}_{\rm sol}[\Delta^+ , \Delta^- , B_1 , B_3] \nonumber\\
 &\equiv&{V}[\Delta^+ , \Delta^- , B_1 , B_3]-V[0 , 0 , 0 , 0] \nonumber\\
 &=&-\int\frac{d^4p}{i(2\pi)^4}2\biggl[
  2\ln\biggl(\frac{F(p,B_1,\Delta)}{[(p_0+\mu)^2-\bar{p}^2]
  [(p_0-\mu)^2-\bar{p}^2]}\biggr) \nonumber\\
 &&\hspace{2.5cm}+\ln\biggl(\frac{F(p,B_3,0)}{[(p_0+\mu)^2-\bar{p}^2]
  [(p_0-\mu)^2-\bar{p}^2]}\biggr)\biggr] \nonumber\\
 &&-\int\frac{d^4p}{i(2\pi)^4}2\biggl[\frac{2}{F(p,B_1,\Delta)}\biggl\{
  [(p_0-\mu)^2-\bar{p}^2-\{B_1(-p)\}^2][(p_0+\mu)^2-\bar{p}^2] \nonumber\\
 &&\hspace{5cm}
  +[(p_0+\mu)^2-\bar{p}^2-\{B_1(p)\}^2][(p_0-\mu)^2-\bar{p}^2] \nonumber\\
 &&\hspace{4cm}-[(p_0)^2-(\bar{p}+\mu)^2]\vert\Delta^-\vert^2
  -[(p_0)^2-(\bar{p}-\mu)^2]\vert\Delta^+\vert^2\biggr\}
  \nonumber\\
 &&\hspace{2cm}+\frac{1}{F(p,B_3,0)}\biggl\{
  [(p_0-\mu)^2-\bar{p}^2-\{B_3(-p)\}^2][(p_0+\mu)^2-\bar{p}^2] \nonumber\\
 &&\hspace{5cm}
  +[(p_0+\mu)^2-\bar{p}^2-\{B_3(p)\}^2][(p_0-\mu)^2-\bar{p}^2]\biggr\}
  \nonumber\\
 &&\hspace{6cm}-6\biggr] \ . 
\label{potm}
\end{eqnarray}
Here $F$ is defined in Eq.~(\ref{Fdef}).
The value of the effective potential in Eq.~(\ref{potm}) is understood as
the energy density of the solution.
Therefore the true vacuum should be determined by evaluating the value of the
effective potential.
The vacuum corresponding to the solution becomes
more stable as the value of $\bar{V}_{\rm sol}$ decreases.

\section{Numerical analysis}
\label{Numerical analysis}
In this section we give the results of our numerical analysis 
for $0.2\leq\mu/\Lambda_{\rm qcd}\leq{1.0}$.
The parameters necessary to carry out the numerical analysis are 
$\Lambda_{\rm qcd}$ and the infrared cutoff parameter $t_f$ 
 in the running coupling.
$\Lambda_{\rm qcd}$ is the unit of the energy scale in our numerical analysis, 
and it is determined by calculating the 
pion decay constant $f_\pi$ for fixed $t_f$ at zero density 
through the Pagels-Stokar formula,\cite{Pag}
\begin{eqnarray}
\label{psf}
 f_\pi^2=\frac{N_c}{4\pi^2}\int{p_E^2}{dp_E^2}
  \frac{B(p)\biggl(B(p)-\frac{p_E^2}{2}\frac{dB(p)}
  {dp_E^2}\biggr)}
  {[p_E^2+B^2(p)]^2} \ .
\end{eqnarray}
We use $f_\pi=88$ MeV in the chiral limit\cite{Gasser:1984yg} as an input.
For $\mu=0$, the dependences of the physical quantities 
on $t_f$ have been shown to be slight around $t_f=0.5$.\cite{Ao}
With this in mind, for the time being, we fix $t_f=0.5$ for general $\mu$. 
{}From these inputs we obtain $\Lambda_{\rm qcd}=604$ MeV. 
Later in this section we study the $t_f$ dependence of our results.

We introduce the framework of our numerical analysis in
 \S\ref{Framework of the numerical analysis}.
Then, we give the solutions for the Majorana masses ($\Delta^-$ and
 $\Delta^+$) fixing $B_{1}=B_{3}=0$ in \S\ref{M}. 
We give the solutions for the Dirac mass ($B_{1}$ and $B_{3}$) with 
$\Delta^-=\Delta^+=0$ in \S\ref{D}.
Finally, in \S\ref{Phase transition},
 we study the phase transition from the hadronic phase 
to the color superconducting phase. 

\subsection{Framework of the numerical analysis}
\label{Framework of the numerical analysis}
In this subsection we summarize the framework of our numerical analysis.
First, as discussed below Eq.~(\ref{diratprop}),
 we note that it is always possible to restrict the $p_4$-integral 
$(p_4=-ip_0)$ to its positive region due to the properties in 
Eqs.~(\ref{majprop}) and (\ref{diratprop}). 
To solve the SDEs numerically, we transform the variables
 $p_4$ and $\bar{p}=\vert\vec{p}\vert$ into new variables $U$ and $X$.
For these transformations, we use density-independent transformations
 in the low density region ($\mu<\mu_0$) and density-dependent
 transformations in the medium density region ($\mu\geq\mu_0$),
 where $\mu_0$ is determined in \S\ref{M}.
In the low density region, where the chiral condensate is formed,
the characteristic scale of the system is $\Lambda_{\rm qcd}$.
The dynamical information comes mainly
from the region $p_4, \ \bar{p}\lsim\Lambda_{\rm qcd}$.
On the other hand, in the high density region,
 where the diquark condensate is formed, the chemical
 potential $\mu$, in addition to $\Lambda_{\rm qcd}$, gives an important scale.
Here, the dynamically important region is 
$p_4\lsim\Lambda_{\rm qcd}$ and
 $\bar{p}\sim\mu$. Therefore we adopt the transformations
 $p_4\rightarrow{U}$ and $\bar{p}\rightarrow{X}$ determined by the following:
\begin{eqnarray}
 \label{des1}
&&p_4=\Lambda_{\rm qcd}\cdot\exp(U) \ ,
 \hspace{0.7cm} \bar{p}=\Lambda_{\rm qcd}\cdot\exp(X) \ 
 \hspace{1.2cm}\mbox{for} \quad \mu<\mu_0 \ ,
 \qquad \\
 \label{des2}
&&p_4=3\mu\cdot\exp(U) \ , \hspace{1cm} \bar{p}=3\mu\cdot\exp(X) \ 
\hspace{1.5cm}\mbox{for} \quad \mu\geq\mu_0 \ . 
 \qquad 
\end{eqnarray}

Under these transformations, 
the integrations over $p_4$ and $\bar{p}$ on the interval $[0,\infty]$
 are converted into integrals over $U$ and $X$ on the interval 
$[-\infty,\infty]$. In the numerical integration, 
we introduced ultraviolet (UV) and infrared (IR) cutoffs for $U$ and $X$,
 restricting their values as
\begin{eqnarray}
 U\in[\Lambda_{IR}, \Lambda_{UV}] \ , \qquad 
  X\in[\lambda_{IR}, \lambda_{UV}] \ .
\end{eqnarray}
 We discretize these intervals of $U$ and $X$ evenly
 into $N_U$ and $N_X$ points, respectively: 
\begin{eqnarray}
 U[I]&=&\Lambda_{IR}+\Delta{U}\cdot{I} , \quad I=0,1,\cdots,N_U-1 , 
  \nonumber\\
 X[J]&=&\lambda_{IR}+\Delta{X}\cdot{J} , \quad J=0,1,\cdots,N_X-1 , 
  \nonumber\\
\end{eqnarray}
where
\begin{eqnarray}
 \Delta{U}=\frac{\Lambda_{UV}-\Lambda_{IR}}{N_U-1} ,\qquad
  \Delta{X}=\frac{\lambda_{UV}-\lambda_{IR}}{N_X-1}. \quad 
\end{eqnarray}
The integrations over $p_4$ and $\bar{p}$ are thus replaced with 
the following summations:
\begin{eqnarray}
 \int{dp_4} \rightarrow \Delta{U}\sum_{I}e^{U[I]}, \qquad
  \int{d\bar{p}} \rightarrow \Delta{X}\sum_{J}e^{X[J]}.
\end{eqnarray}
In the present analysis, for the UV and IR cutoffs, we use 
\begin{eqnarray}
p_4: \ (\Lambda_{IR},\Lambda_{UV})=(-12.5,2.5) \ , \quad 
\bar{p}: \ (\lambda_{IR},\lambda_{UV})=(-3.5,2.5) \ .
\label{range}
\end{eqnarray}
The validity of these choices is checked below.

We solved the SDEs with an iteration method.
Starting from a set of trial functions, we updated the mass functions
 with the SDE:
\begin{eqnarray}
 \left\{
  B_{1,3{\rm old}} \,,\, \Delta^{\pm}_{\rm old}
\right\}
\Rightarrow
\fbox{Right-hand sides of SDE's (3.6)-(3.9)}
\Rightarrow
\left\{
  B_{1,3{\rm new}} \,,\, \Delta^{\pm}_{\rm new}
\right\}
\ . \nonumber\\
\end{eqnarray}
Then we stopped the iteration when the convergence condition
\begin{eqnarray}
 &&\varepsilon\Lambda_{\rm qcd}^6>\int\frac{d^4p}{(2\pi)^4}
  \frac{1}{4}{\rm tr}
  \biggl[\biggl(\frac{\delta{V}}{\delta[S_F(p)]}\biggr)^\dag
  \biggl(\frac{\delta{V}}{\delta[S_F(p)]}\biggr)\biggr] \nonumber\\
 &&\hspace{2cm}=\int\frac{d^4p}{(2\pi)^4}\biggl\{
  2\vert{B}_{1\rm old}(p)-B_{1\rm new}(p)\vert^2
  +\vert{B}_{3\rm old}(p)-B_{3\rm new}(p)\vert^2
  \nonumber\\
 &&\hspace{3cm}+2\vert\Delta^+_{\rm old}(p)-\Delta^+_{\rm new}(p)\vert^2
  +2\vert\Delta^-_{\rm old}(p)-\Delta^-_{\rm new}(p)\vert^2
  \biggr\}
\end{eqnarray}
was satisfied, with suitably small $\varepsilon$.
 In the present analysis we set $\varepsilon=10^{-10}$. 

\subsection{$B=0$ , $\Delta\neq{0}$ solution}
\label{M}
In this subsection we report the results of the numerical solution of
 the SDEs with Dirac masses fixed to zero ($B_1=B_3=0$).
The initial trial functions used here are
\begin{eqnarray}
&& B_1(p) = B_3(p) = 0 \ , \nonumber\\
&& \Delta^{+}(p) = 0 \ , \nonumber\\
&& \Delta^{-}(p) = \Lambda_{\rm qcd} \ .
\end{eqnarray}
In this case, the outputs become $B_{1}(p), B_{3}(p)=0$ and $\Delta^-(p),
 \Delta^+(p)\neq{0}$ for all $\mu$.
We call this solution the color symmetry breaking
(CSB) solution.   

To check the validity of the UV and IR cutoffs in Eq.~(\ref{range})
 we plot the solutions $\Delta^-(p)$ and $\Delta^+(p)$ and 
the integrand of $\bar{V}_{\rm sol}$
 at $\mu/\Lambda_{\rm qcd}=0.70$ in Figs.~2 and 3, respectively.
Figure~2 shows that both $\Delta^-$ and $\Delta^+$ become small
 in the UV region of $p_4$, as well as that of $\bar{p}$.
It has been shown(e.g., in Ref.~\citen{Hong}) that
$\Delta^+$ is very small compared with $\Delta^-$ in the high density region.
In the medium density region,
Fig.~2 shows that $\Delta^+$ is of the same order as
$\Delta^-$, but their momentum dependences are rather different.
This implies that it is important to include two Majonara masses
$\Delta^+$ and $\Delta^-$ in the low density and intermediate density
regions.
{}From Fig.~3 we see that the dominant contribution to the effective
potential lies within the integration range. 
These figures show that the choices of the ranges in
 Eq.~(\ref{range}) are sufficient at $\mu/\Lambda_{\rm qcd}=0.70$.
We carried out similar analyses for all the cases we study in this
paper, and confirmed that the choices of the ranges in Eq.~(\ref{range}) are
 sufficient for the present purposes.

\vspace{0.3cm}

\begin{center}
\begin{minipage}{6cm}
 \epsfxsize=6cm
 \epsfbox{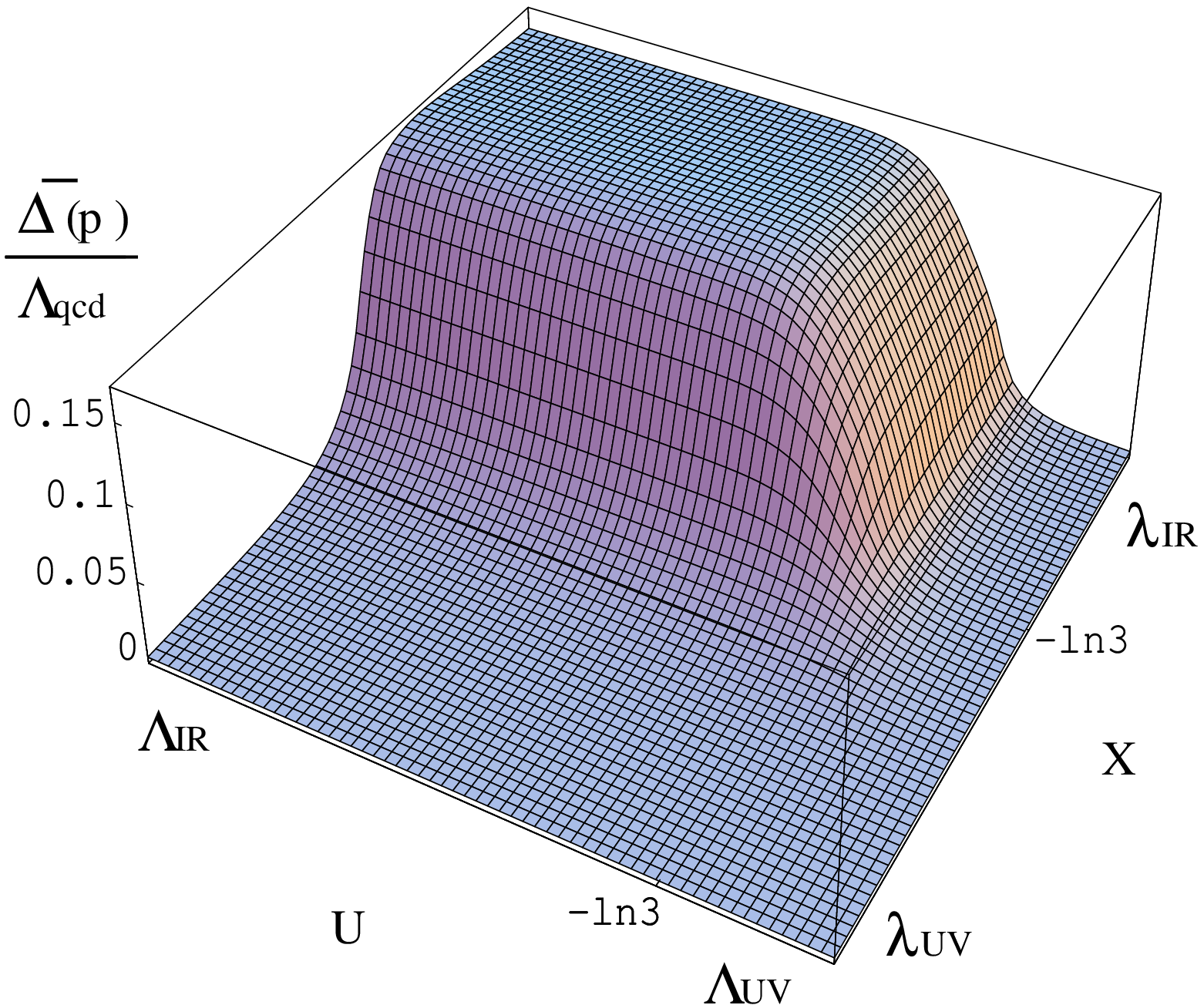}
\end{minipage}\hspace{1cm}
\begin{minipage}{6cm}
 \epsfxsize=6cm
 \epsfbox{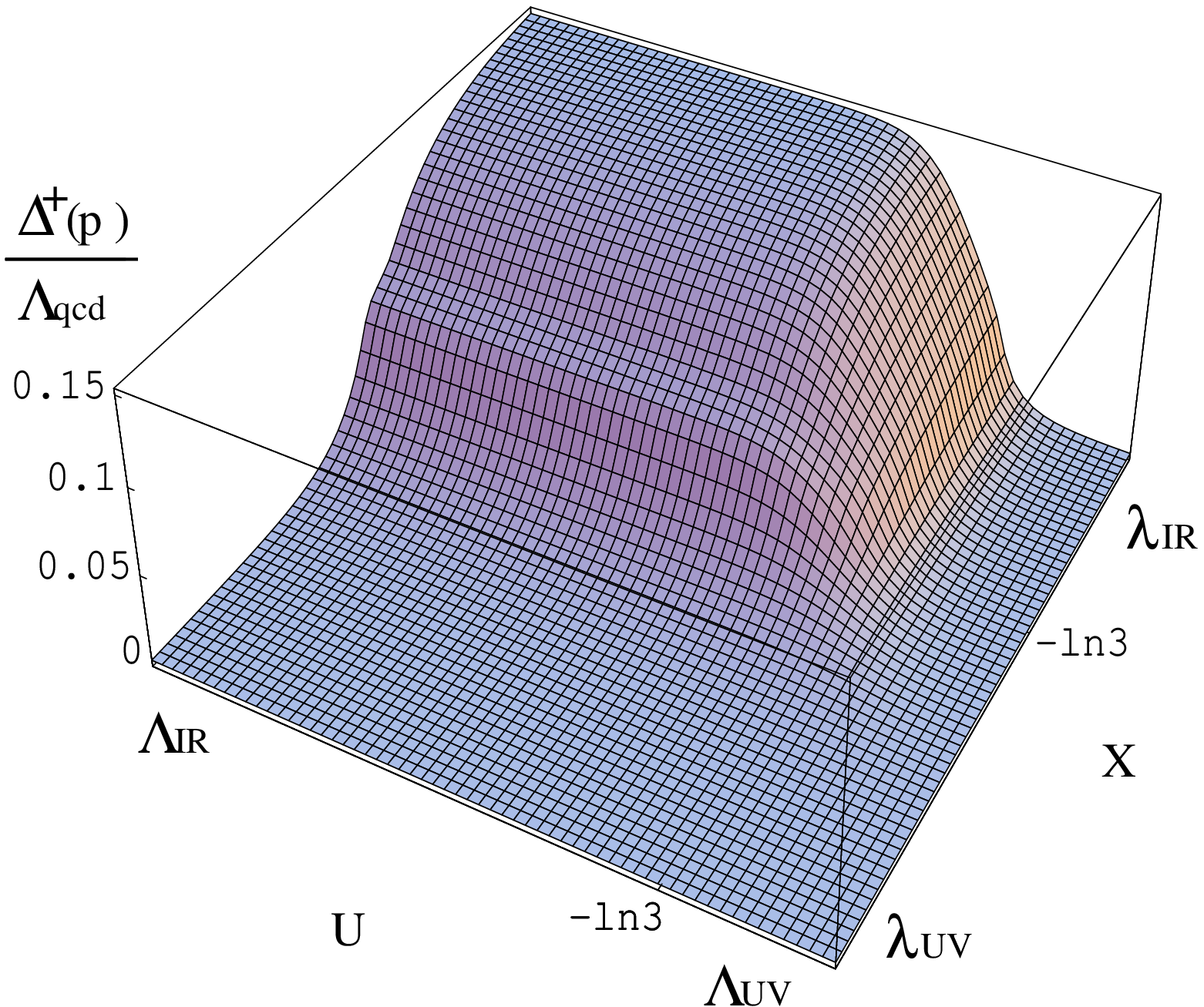}
\end{minipage}
\end{center}

\vspace{0.3cm}

\begin{center}
\begin{minipage}{12cm}
 Fig.~2$\cdot$ Momentum dependences of Majorana mass functions 
 $\Delta^-(p)/\Lambda_{\rm qcd}$ and $\Delta^+(p)/\Lambda_{\rm qcd}$
 at $\mu/\Lambda_{\rm qcd}=0.70$.
 Here $U=\ln(p_4/3\mu)$ and $X=\ln(\bar{p}/3\mu)$.
\end{minipage}
\end{center}

\vspace{0.3cm}

\begin{center}
\begin{minipage}{10cm}
 \epsfxsize=10cm
 \epsfbox{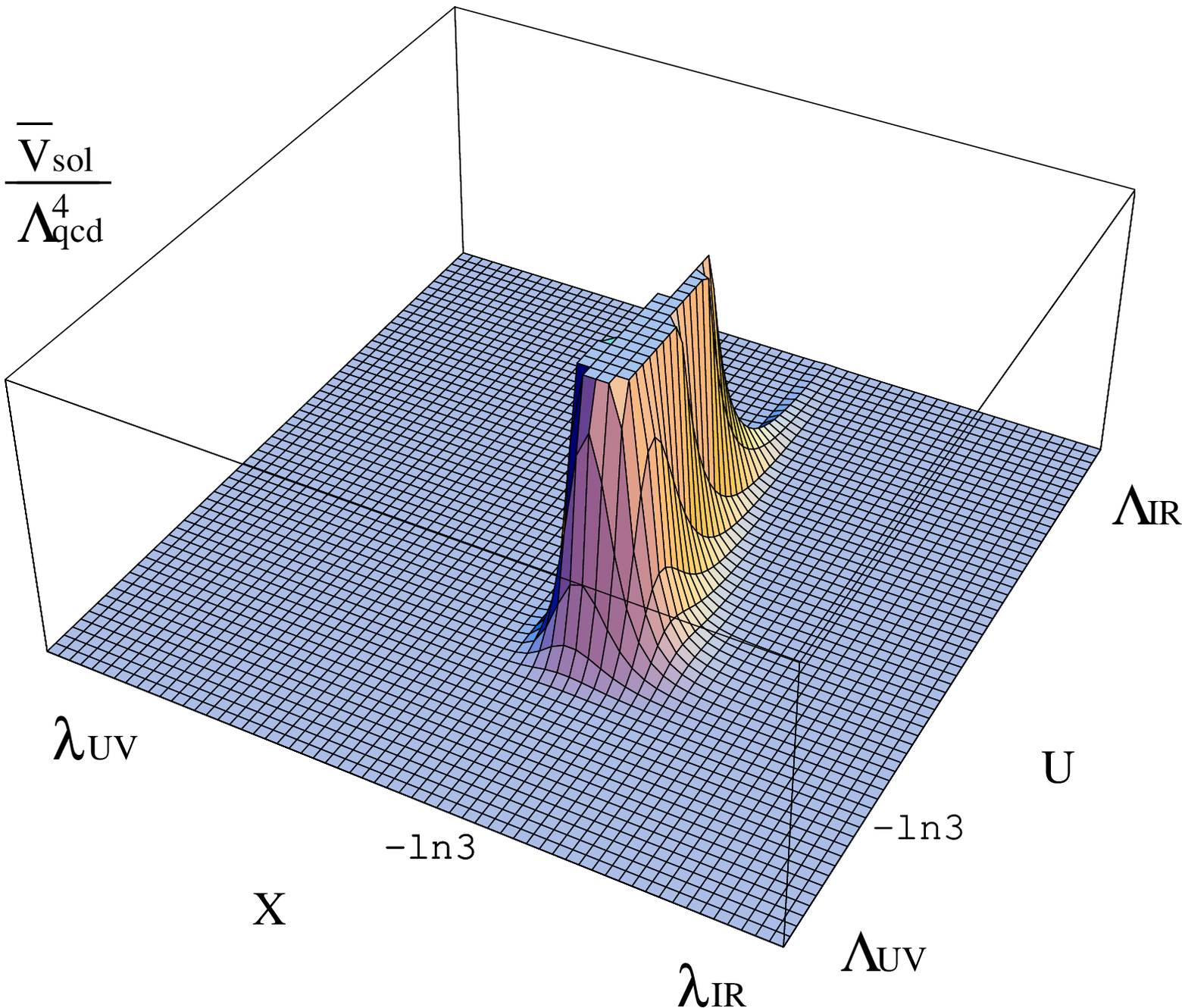}
\end{minipage}
\end{center}
\begin{center}
\begin{minipage}{12cm}
 Fig.~3$\cdot$ Integrand of $\bar{V}_{\rm sol}/\Lambda_{\rm qcd}^4$ 
 at $\mu/\Lambda_{\rm qcd}=0.70$. 
 The upper 9/10 of the figure is clipped.
 Here $U=\ln(p_4/3\mu)$ and $X=\ln(\bar{p}/3\mu)$.
\end{minipage}
\end{center}

\vspace{1cm}

\begin{center}
\begin{minipage}{12cm}
 \epsfxsize=12cm
 \epsfbox{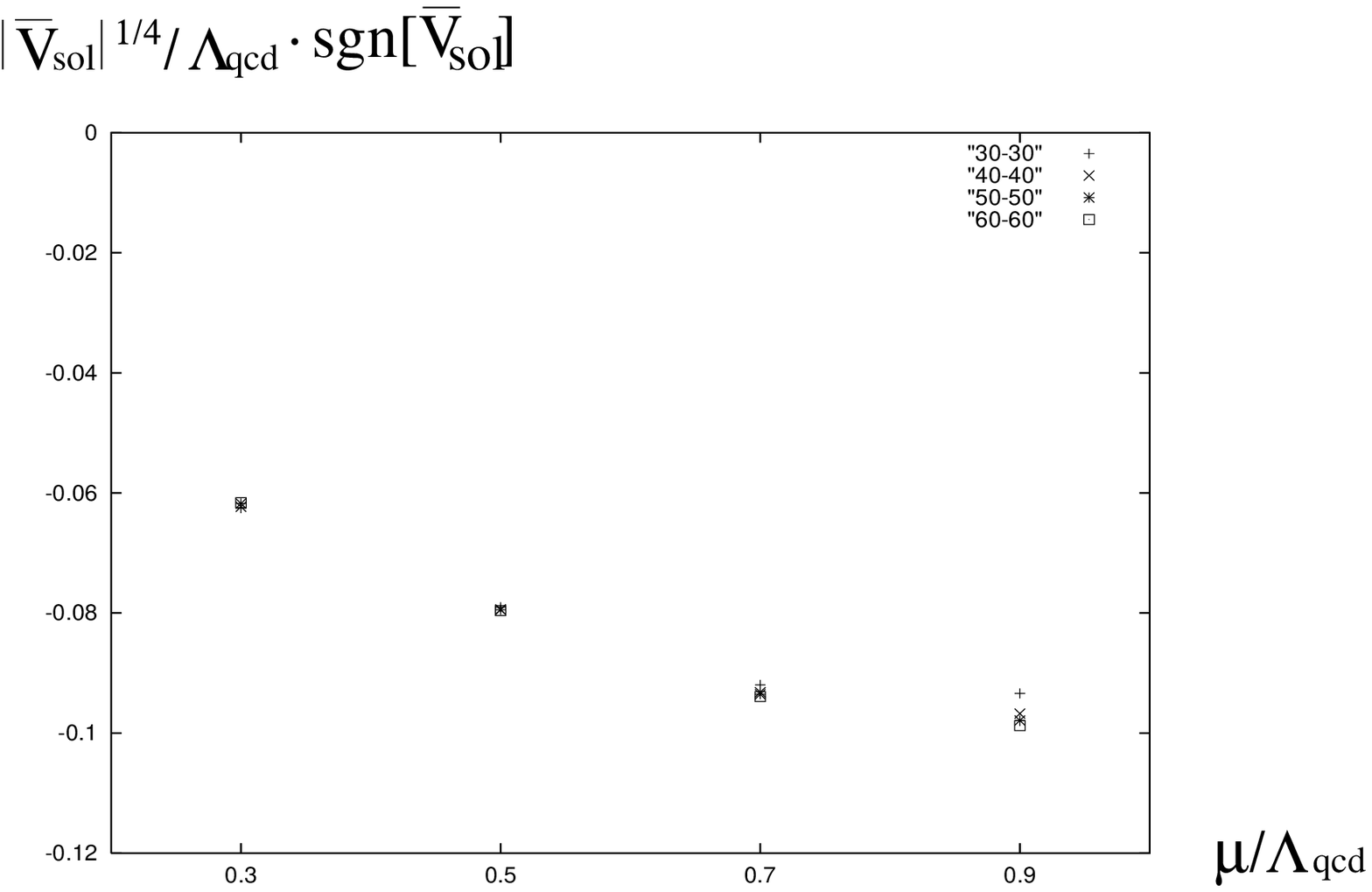}
 Fig.~4$\cdot$ Typical values of
 $\vert\bar{V}_{\rm sol}\vert^{1/4}/{\Lambda_{\rm qcd}}\cdot{\rm sgn[\bar{V}_{\rm sol}]}$ for four choices of the fineness of the discretization, 
 ($N_U, N_X$)=(30,30), (40,40), (50,50) and (60,60).
\end{minipage}
\end{center}

\vspace{1cm}

\begin{center}
\begin{minipage}{12cm}
 \epsfxsize=12cm
 \epsfbox{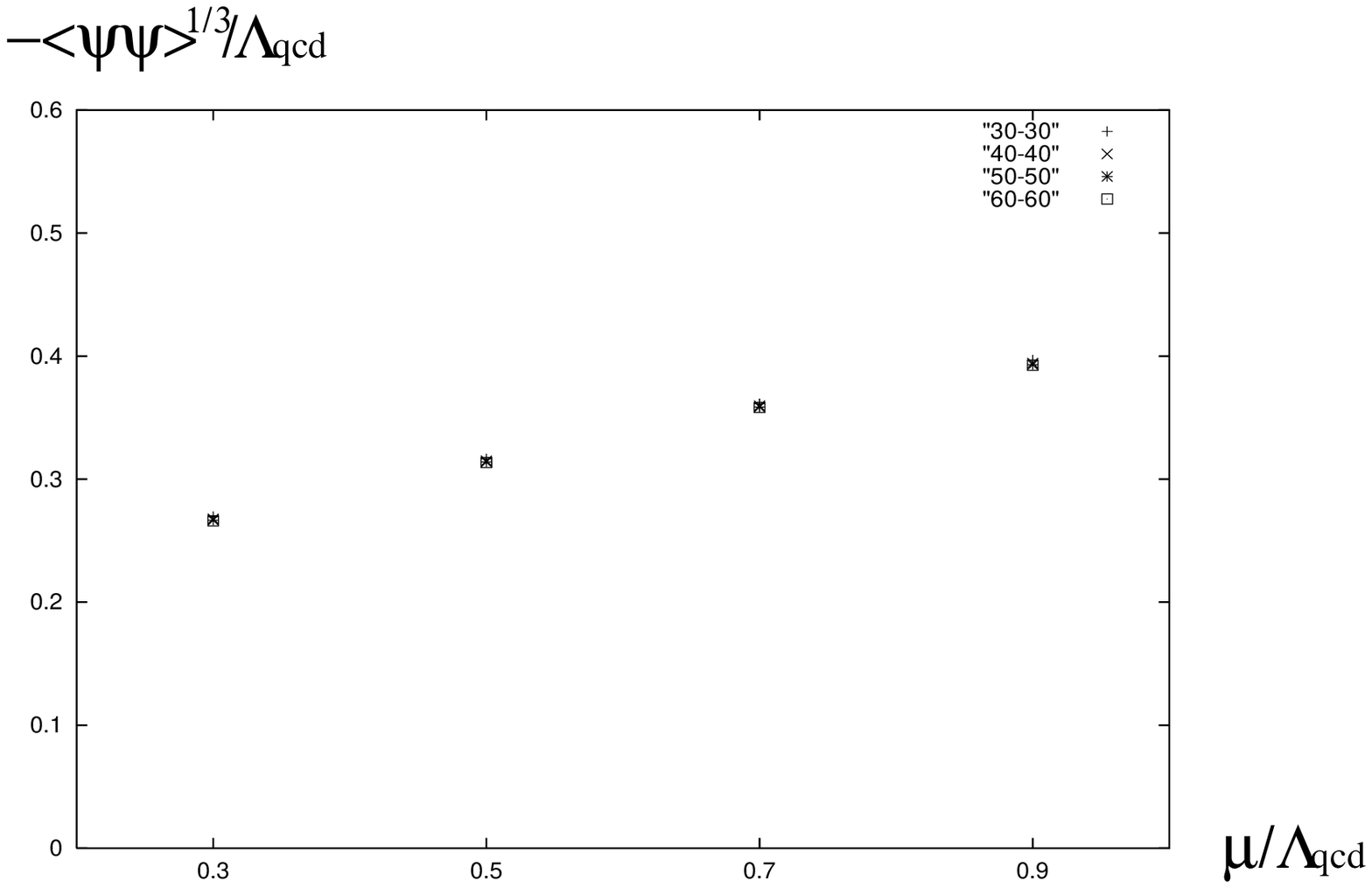}
 Fig.~5$\cdot$ Typical values of $-\langle\psi\psi\rangle^{1/3}_{\rm 1GeV}
 /{\Lambda_{\rm qcd}}$ for four choices of the fineness of the discritization, 
 $(N_U,N_X)$=$(30,30)$, $(40,40)$, $(50,50)$ and $(60,60)$.
\end{minipage}
\end{center}

\vspace{1cm}

\begin{center}
\begin{minipage}{12cm}
 \epsfxsize=12cm
 \epsfbox{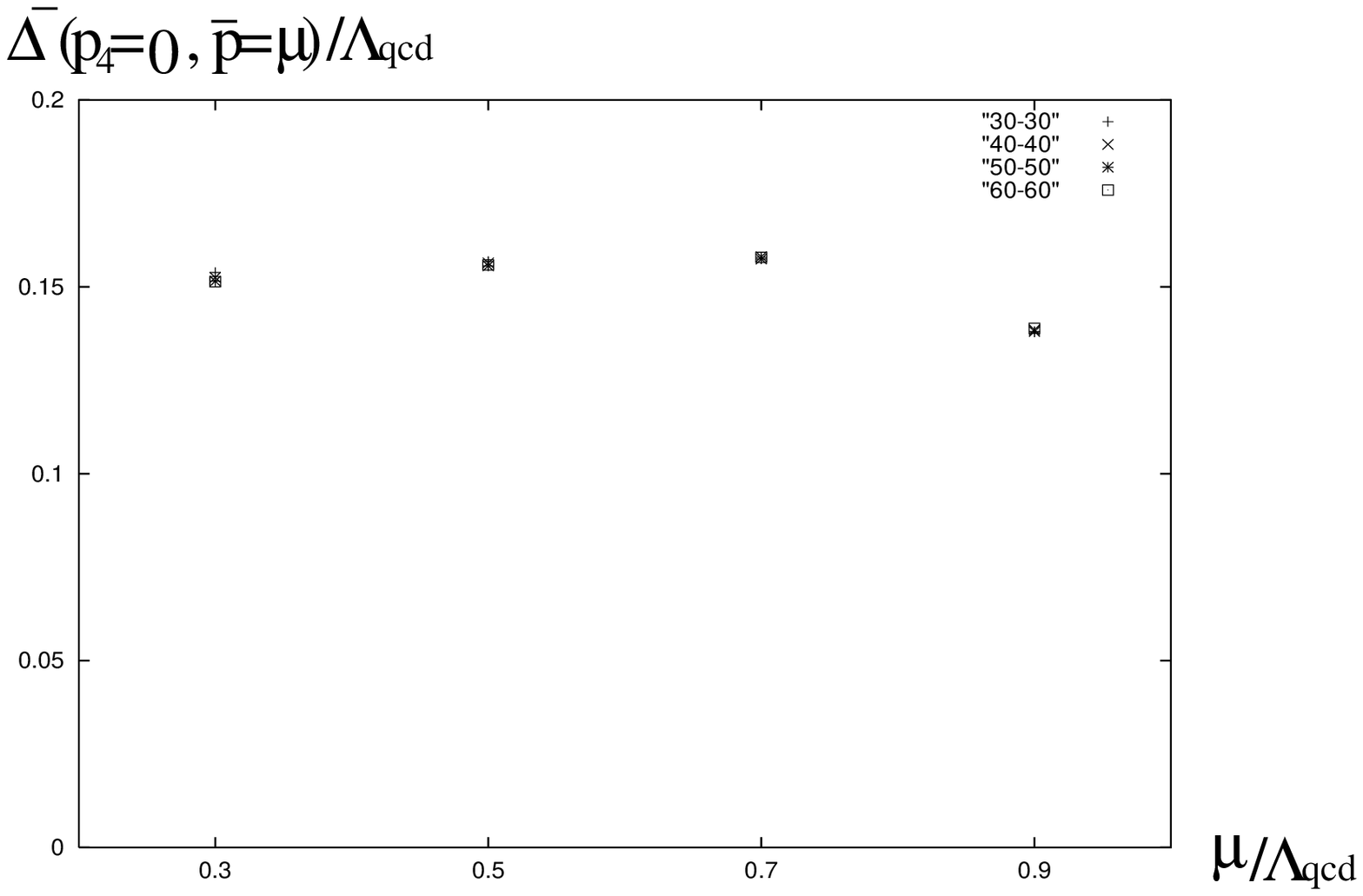}
 Fig.~6$\cdot$ Typical values of 
 $\Delta^-(p_4=0,\bar{p}=\mu)/\Lambda_{\rm qcd}$
 for four choices of the fineness of the discretization, 
 ($N_U, N_X$)=(30,30), (40,40), (50,50) and (60,60).
\end{minipage}
\end{center}

\vspace{1cm}

\begin{center}
\begin{minipage}{12cm}
 \epsfxsize=12cm
 \epsfbox{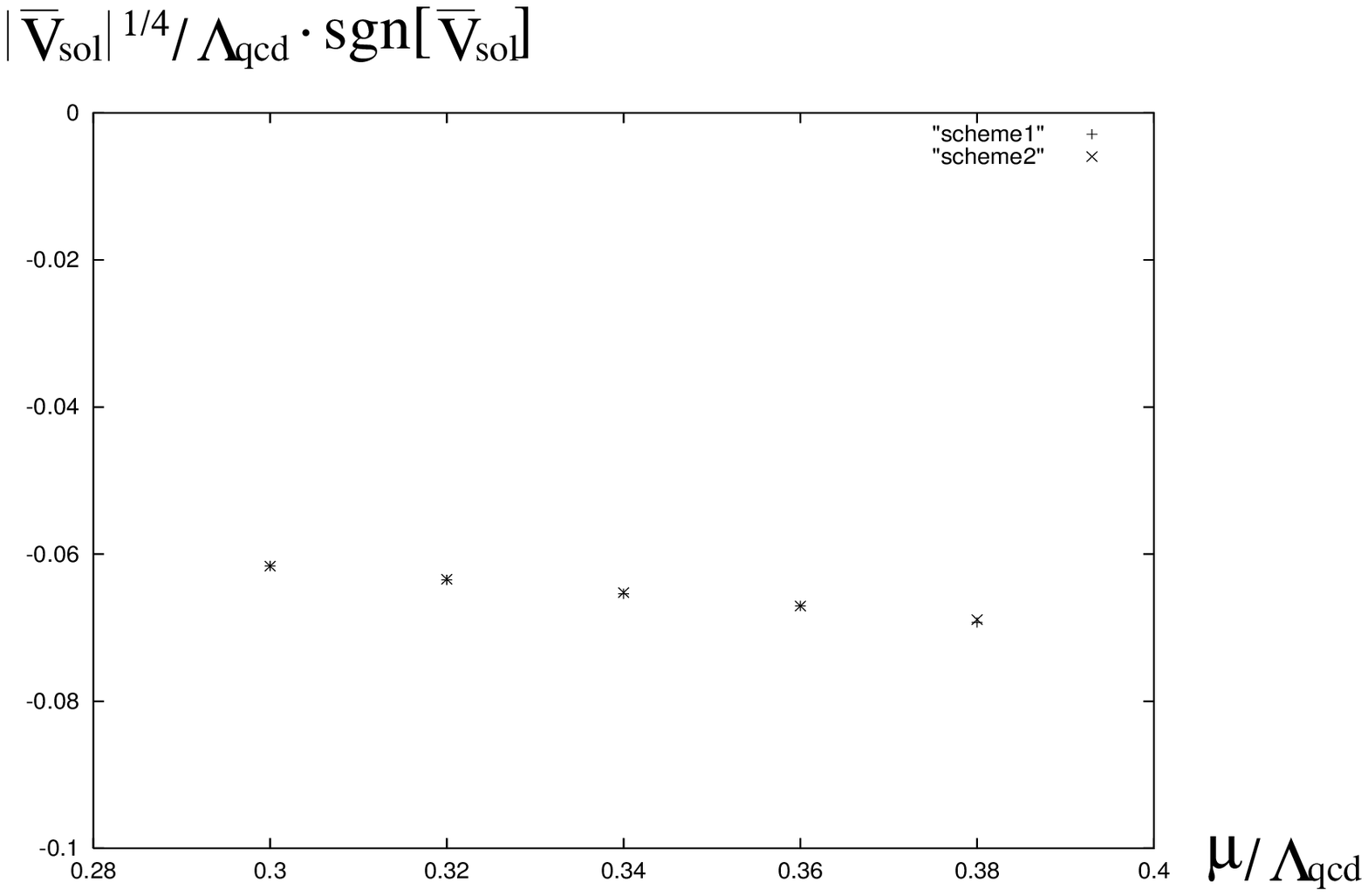}
 Fig.~7$\cdot$ Typical values of
 $\vert\bar{V}_{\rm sol}\vert^{1/4}/{\Lambda_{\rm qcd}}\cdot{\rm sgn[\bar{V}_{\rm sol}]}$ in two different schemes of the discritization for 
 $0.30\leq\mu/\Lambda_{\rm qcd}\leq0.38$.
Here, ``scheme 1'' is that in which Eq.~(\ref{des1}) is used and
 ``scheme 2'' is that in which Eq.~(\ref{des2}) is used.
\end{minipage}
\end{center}

\vspace{1cm}

\begin{center}
\begin{minipage}{12cm}
 \epsfxsize=12cm
 \epsfbox{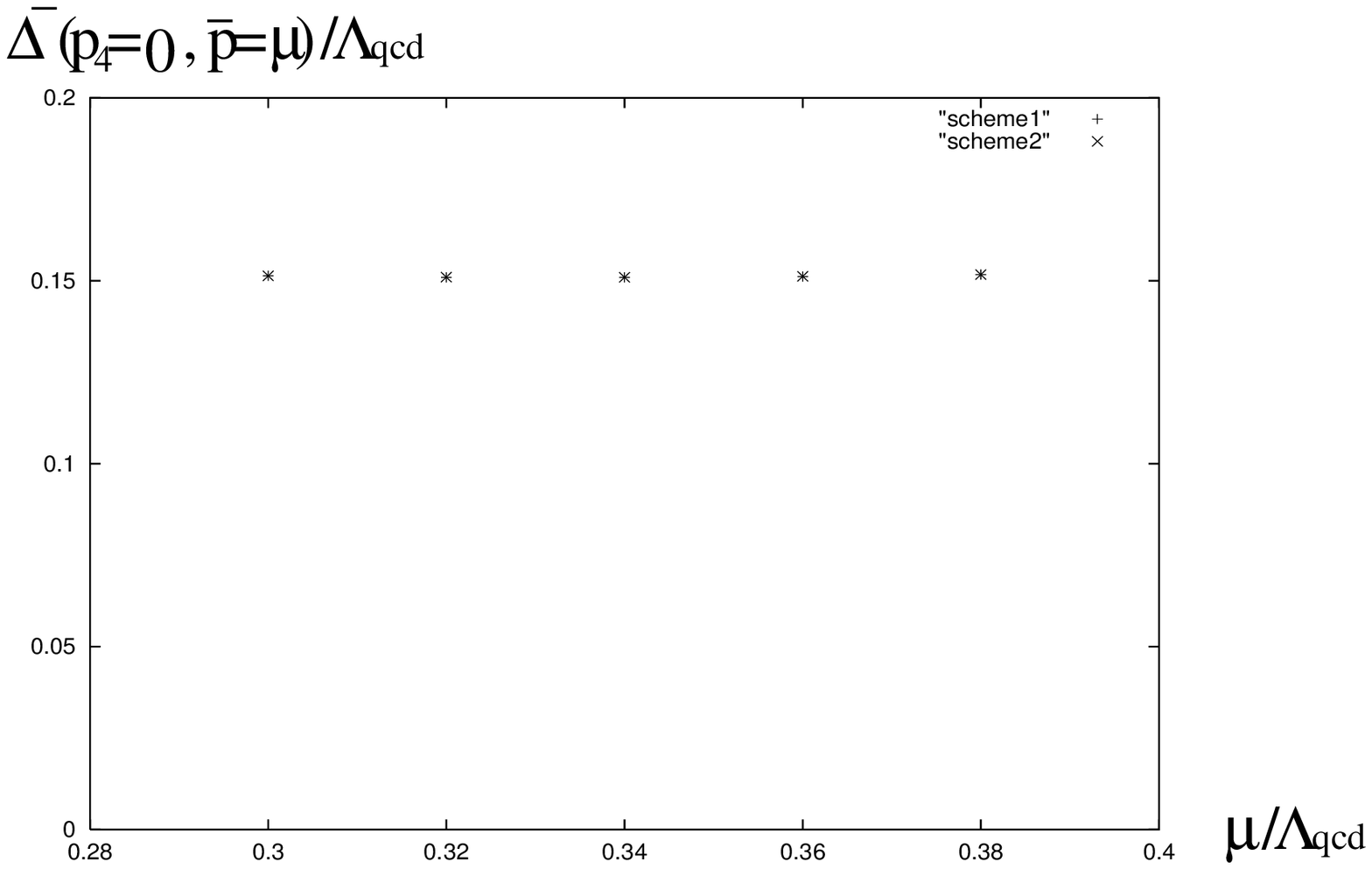}
 Fig.~8$\cdot$ Typical values of 
 $\Delta^-(p_4=0,\bar{p}=\mu)/\Lambda_{\rm qcd}$
 in two different schemes of the descritization for 
 $0.30\leq\mu/\Lambda_{\rm qcd}\leq0.38$.
\end{minipage}
\end{center}

\vspace{1cm}

Next, we elucidate the dependence of the results on the size of discretization.
We show typical values of 
$\vert\bar{V}_{\rm sol}\vert^{1/4}/{\Lambda_{\rm qcd}}\cdot{\rm
sgn[\bar{V}_{\rm sol}]}$\footnote{sgn$[\bar{V}_{\rm sol}]=1$ for
$\bar{V}_{\rm sol}\geq{0}$ and $-1$ for $\bar{V}_{\rm sol}<{0}$.}
 in Fig.~4, 
$-\langle\psi\psi\rangle^{1/3}_{\rm 1GeV}/{\Lambda_{\rm qcd}}$ in Fig.~5 
and $\Delta^-(p_4=0,\bar{p}=\mu)/\Lambda_{\rm qcd}$
 in Fig.~6 for four choices 
of the fineness of the discretization, 
$(N_U,N_X)$=(30,30), (40,40), (50,50) and (60,60). 
To obtain $\langle\psi\psi\rangle_{\rm 1GeV}$, we used
Eq.~(\ref{qqrenom}) with 
\begin{eqnarray}
&&\Lambda^2=\Lambda_{qcd}^2[\exp(2\Lambda_{UV})
+\exp(2\lambda_{UV})] \ 
\hspace{1.3cm}\mbox{for} \quad \mu<\mu_0 \ ,
 \qquad \\
 &&\Lambda^2=(3\mu)^2[\exp(2\Lambda_{UV})
  +\exp(2\lambda_{UV})] \  
  \hspace{1.1cm}\mbox{for} \quad \mu\geq\mu_0 \ . 
  \qquad 
\end{eqnarray}
These figures show that the values $N_U=60$ and $N_X=60$ are
sufficiently large for the present purpose.

Let us next check that the two schemes of the discritization described by
Eqs.~(\ref{des1}) and (\ref{des2}) are smoothly connected with each
other. For this purpose, we calculated the values 
of $\bar{V}_{\rm sol}$ and $\Delta^-(p_4=0,\bar{p}=\mu)$
for several values of $\mu$ in
 $0.30\leq\mu/\Lambda_{\rm qcd}\leq{0.38}$ using the two schemes.
We show the results in Fig.~7 and Fig.~8.
These clearly show that the two schemes are smoothly connected with each
other around $\mu_0/\Lambda_{\rm qcd}=1/3$. 
For this reason, in the following analysis we fix 
$\mu_0/\Lambda_{\rm qcd}=1/3$.

\vspace{1cm}

\begin{center}
\begin{minipage}{12cm}
 \epsfxsize=12cm
 \epsfbox{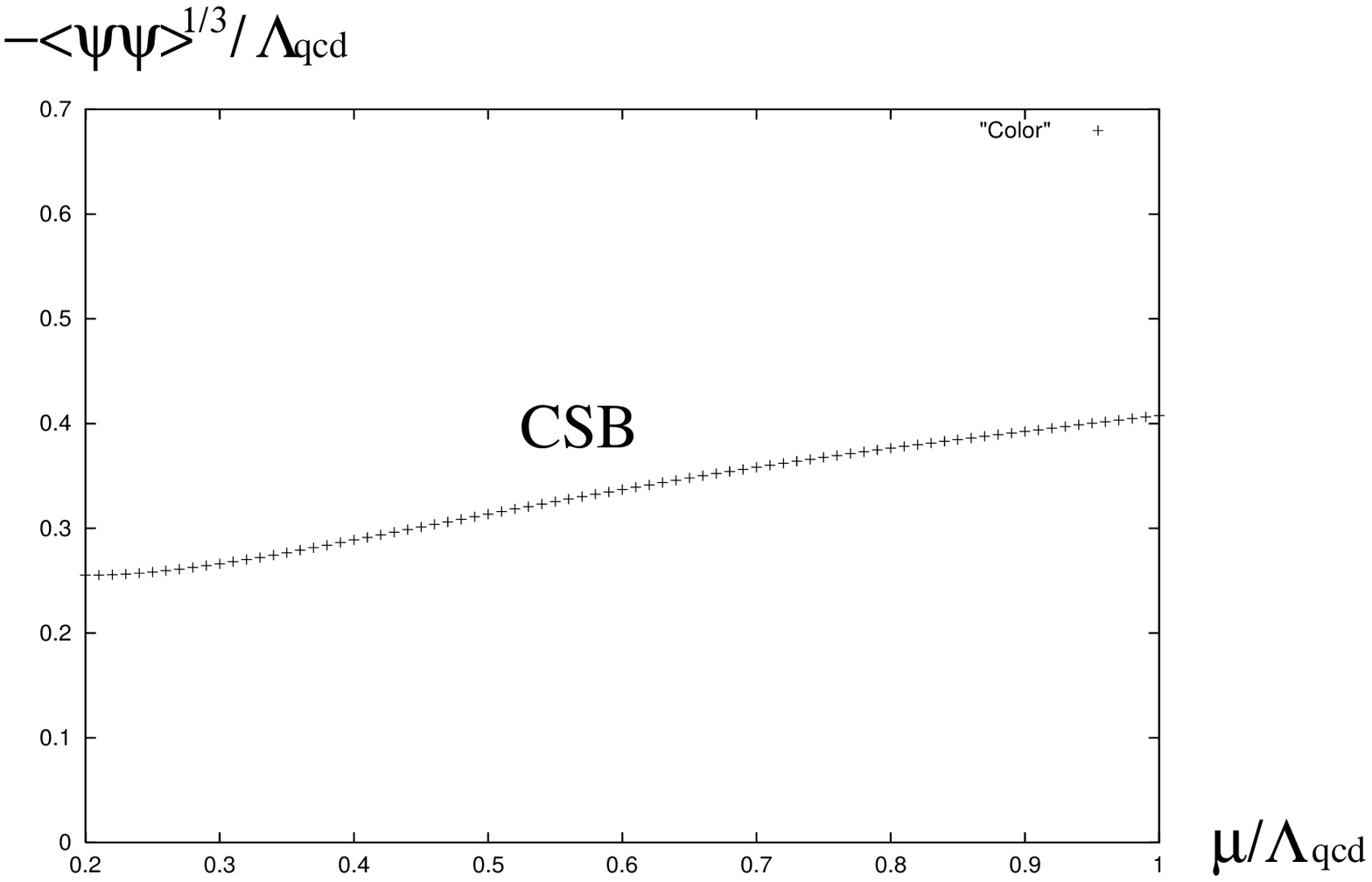}
 Fig.~9$\cdot$ Chemical potential dependence
 of $-\langle\psi\psi\rangle^{1/3}_{\rm 1GeV}/{\Lambda_{\rm qcd}}$ \\ for
 $0.2\leq\mu/\Lambda_{\rm qcd}\leq{1.0}$. 
\end{minipage}
\end{center}

\vspace{1.5cm}

\begin{center}
\begin{minipage}{12cm}
 \epsfxsize=12cm
 \epsfbox{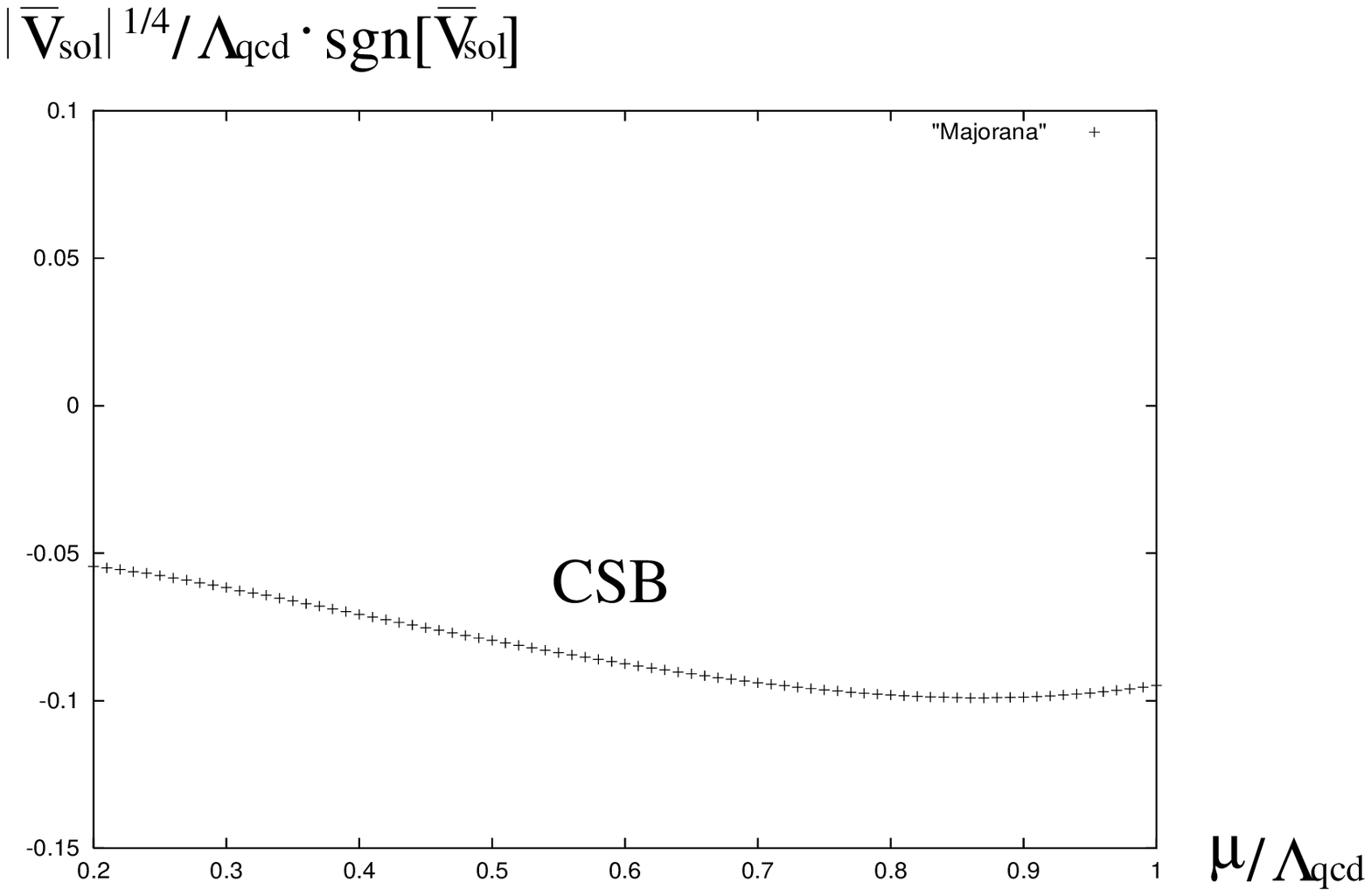}
 Fig.~10$\cdot$ Chemical potential dependence
 of $\vert\bar{V}_{\rm sol}\vert^{1/4}/{\Lambda_{\rm qcd}}
 \cdot{\rm sgn[\bar{V}_{\rm sol}]}$ \\ for
 $0.2\leq\mu/\Lambda_{\rm qcd}\leq{1.0}$.
\end{minipage}
\end{center}

\vspace{1cm}

\newpage

Now, we display the resultant values of the diquark condensate in Fig.~9,
where we have used Eq.~(\ref{qqrenom}) to obtain 
$\langle \psi \psi \rangle_{\rm 1GeV}$.
Figure~9 shows that $\langle \psi \psi \rangle_{\rm 1GeV}$ is between 
$- ( 250\, \mbox{MeV})^3$ and $- ( 150\, \mbox{MeV})^3$,
which is comparable to the value of the chiral condensate for $\mu=0$:
$\langle \bar{\psi} \psi \rangle_{\rm 1GeV} 
= -(225 \pm 25\,\mbox{MeV})^3$.~\cite{Gas}
Furthermore, this figure shows that
there exists a diquark condensate for all $\mu$ we studied.

As discussed in \S\ref{Effective potential and
 Schwinger-Dyson equation},
the true vacuum is determined by evaluating the value of the effective
potential at the solution.
We plot the chemical potential dependence of $\vert\bar{V}_{\rm
sol}\vert^{1/4}/{\Lambda_{\rm qcd}}\cdot{\rm sgn[\bar{V}_{\rm sol}]}$
 in Fig.~10. This figure shows that 
the value of the effective potential is always negative, which implies
that the CSB vacuum is always more stable than the trivial 
vacuum with $B=\Delta=0$.
This is consistent with the result obtained in Ref.~\citen{Rober},
where the SDE was converted into an algebraic equation by using the
confining model gluon propagator.

\vspace{1cm}

\begin{center}
\begin{minipage}{12cm}
 \epsfxsize=12cm
 \epsfbox{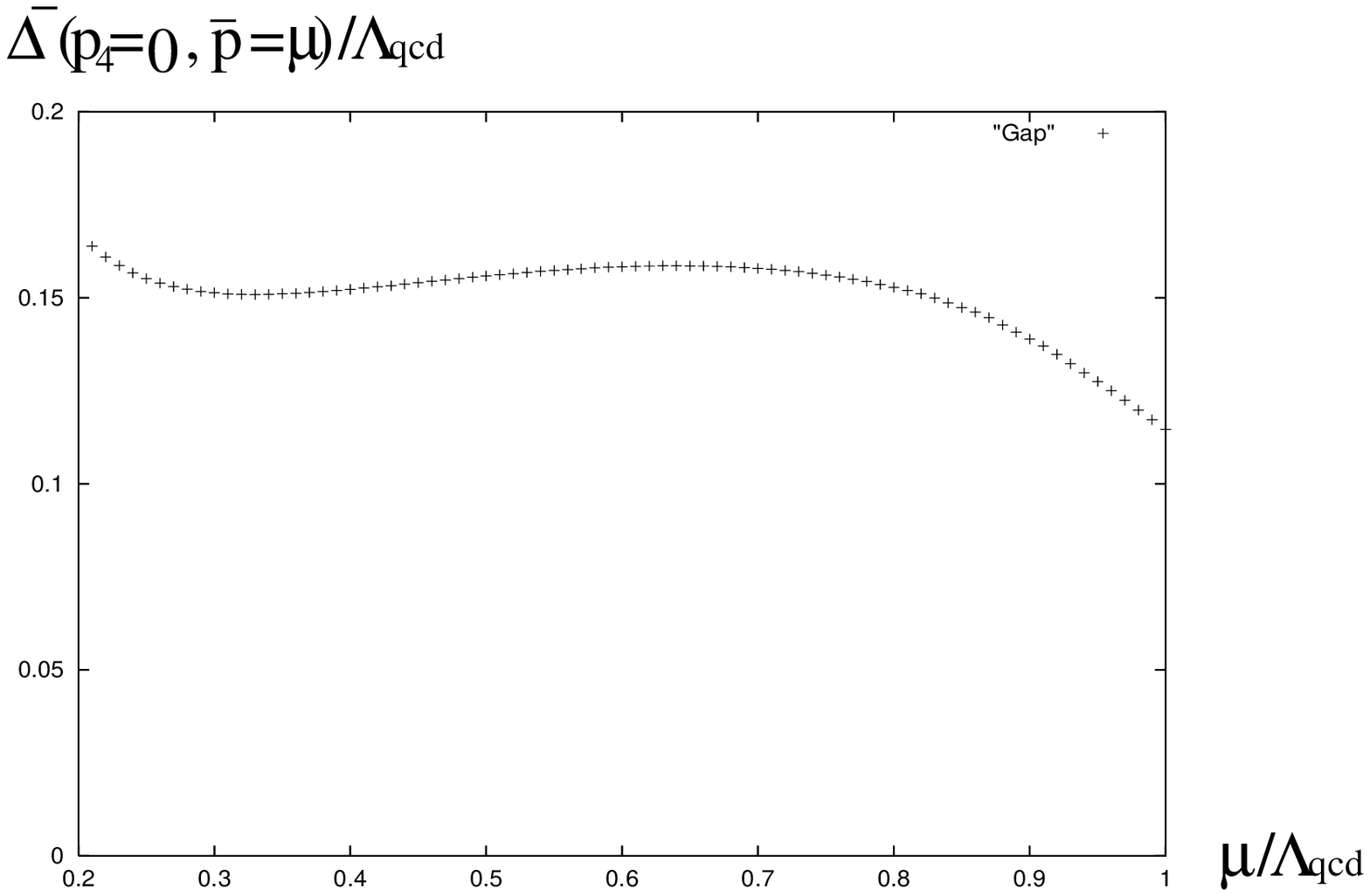}
 Fig.~11$\cdot$ Chemical potential dependence
 of $\Delta^-(p_4=0,\bar{p}=\mu)/\Lambda_{\rm qcd}$ for
 $0.2\leq\mu/\Lambda_{\rm qcd}\leq{1.0}$.
\end{minipage}
\end{center}

\vspace{1cm}

Next, we plot the chemical potential dependence of 
$\Delta^-(p_4=0,\bar{p}=\mu)/\Lambda_{\rm qcd}$ in Fig.~11.
This shows that the Majonara mass gap is on the order of 100 MeV
 in this low density region.
 We also plot the $\bar{p}$ dependence of
$\Delta^-(p_4=0,\bar{p})/\Lambda_{\rm qcd}$ at
$\mu/\Lambda_{\rm qcd}=0.5\mbox{ and }1.0$ in Fig.~12
and the $p_4$ dependence of $\Delta^-(p_4,\bar{p}=\mu)/\Lambda_{\rm
qcd}$ in Fig.~13.
These figures show that the entire scale of the mass function 
becomes small when we increase $\mu$. 
{}From Fig.~12, we see that $\Delta^-$ has a peak around
 $\bar{p}=\mu$ ($X=-\ln3\cong{-1.1}$). However this peak is not sharp, 
and this implies that the effects of the quarks away from the Fermi
surface are important in the medium density region.

\vspace{1cm}

\begin{center}
 \epsfxsize=10cm
\ \epsfbox{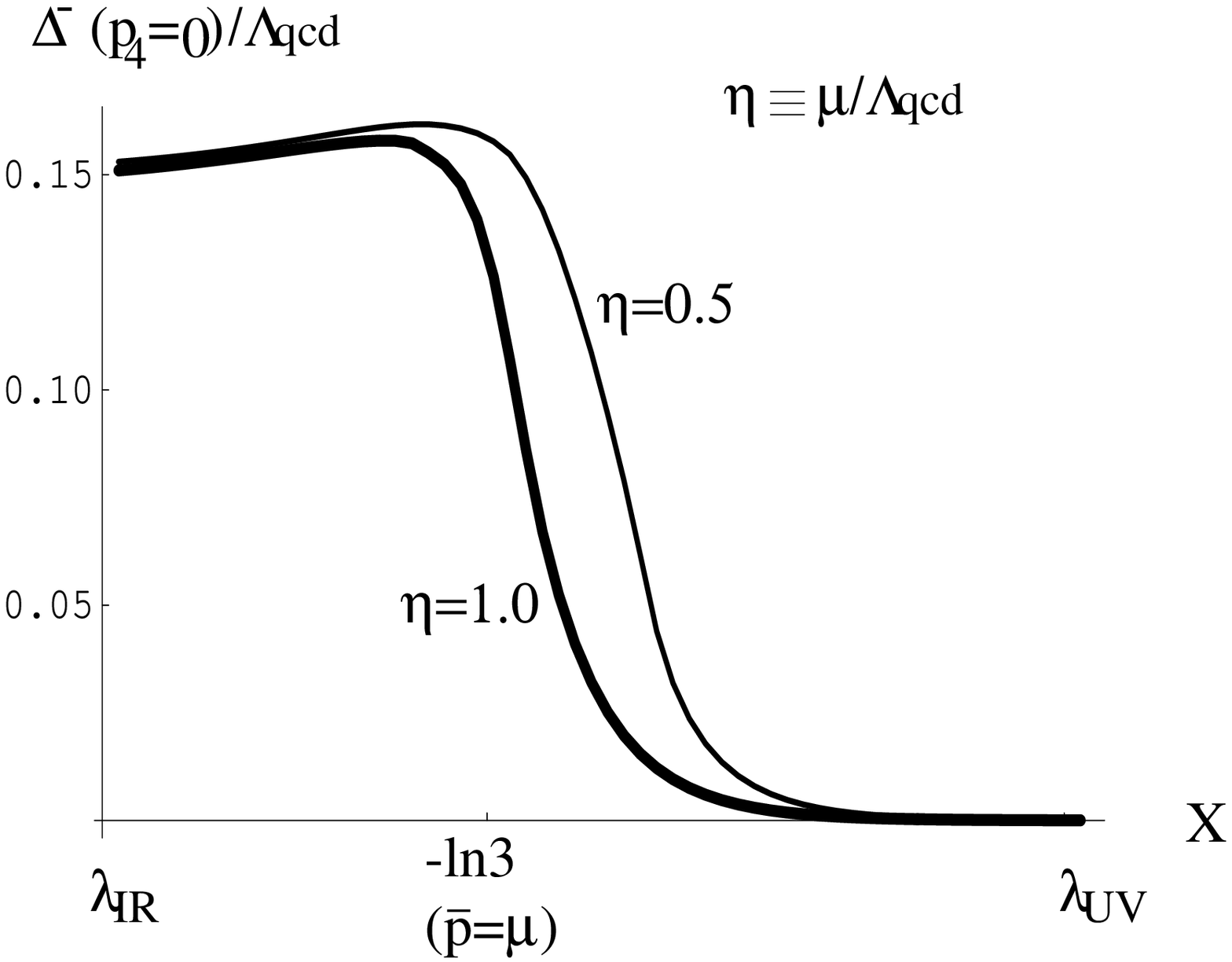}
\\ 
\begin{minipage}{12cm}
Fig.~12$\cdot$ $\bar{p}$ dependence of $\Delta^-(p)/\Lambda_{\rm qcd}$ at
 $p_4=0$ \\ for $\eta\equiv\mu/\Lambda_{\rm qcd}=0.5\mbox{ and }1.0$. 
 Here $X = \ln ( \bar{p}/{3\mu} )$.
\end{minipage}
\end{center}

\vspace{1.5cm}

\begin{center}
 \epsfxsize=10cm
\ \epsfbox{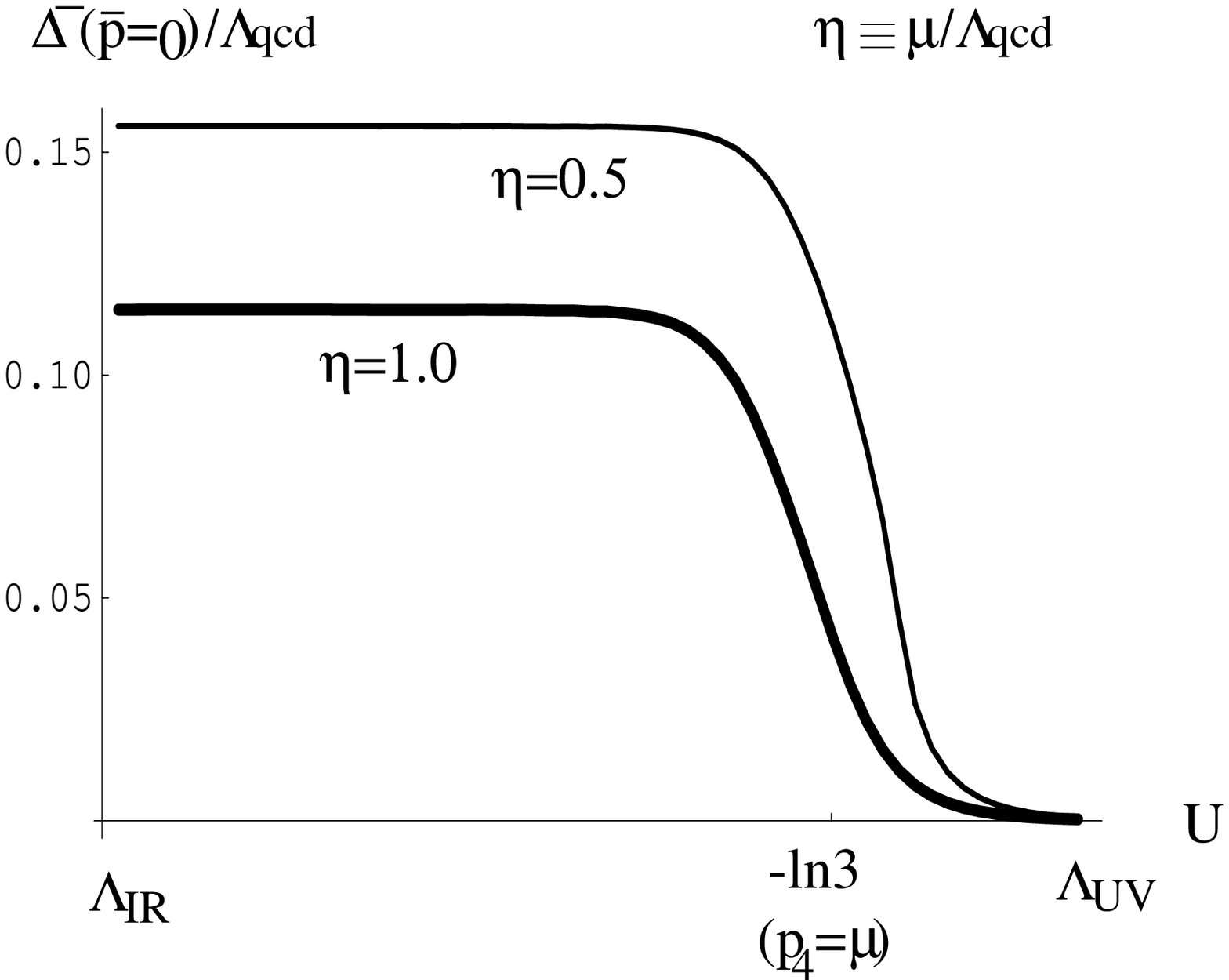}
\\
\begin{minipage}{12cm}
Fig.~13$\cdot$ $p_4$ dependence of $\Delta^-(p)/\Lambda_{\rm qcd}$ 
at $\bar{p}=\mu$ \\ for $\eta\equiv\mu/\Lambda_{\rm qcd}=0.5\mbox{ and }1.0$.
 Here $U = \ln (p_4/{3\mu} )$.
\end{minipage} 
\end{center}

\vspace{1cm}

\newpage

\subsection{$B\neq{0}$ , $\Delta=0$ solution}
\label{D}
In this subsection we report the results of the numerical solution of
the SDEs with the Majorana masses fixed to zero ($\Delta^\pm(p)=0$).
The initial trial functions used here are
\begin{eqnarray}
&& B_1(p) = \Lambda_{\rm qcd} \ , \nonumber\\
&& B_3(p) = 0 \ , \nonumber\\
&& \Delta^{+}(p) = \Delta^{-}(p) = 0 \ .
\end{eqnarray}
In this case, the outputs become $B_1(p)=B_3(p)$, $\Delta^\pm(p)=0$ for 
all $\mu$. We obtain $B_{1,3}(p)\neq{0}$ for small $\mu$, and
 $B_{1,3}(p)={0}$ for large $\mu$.
We call the former solution the
 chiral symmetry breaking ($\chi$SB) solution.

In Fig.~14 we plot the solutions ${\rm Re}[B_1(p)]$ and ${\rm Im}[B_1(p)]$
(or equivalently ${\rm Re}[B_3(p)]$ and ${\rm Im}[B_3(p)]$) at
$\mu/\Lambda_{\rm qcd}=0.30$.
This figure shows that both the real part and the imaginary part
 become small above $\Lambda_{\rm qcd}$ ($U$, $X$=0),
 and that the imaginary part has a peak around $p_4=\Lambda_{\rm
qcd}$ ($U$=0). These structures are consistent with those obtained in Ref.~\citen{Ha}.

\vspace{0.5cm}

\begin{center}
\begin{minipage}{6cm}
 \epsfxsize=6cm
\ \epsfbox{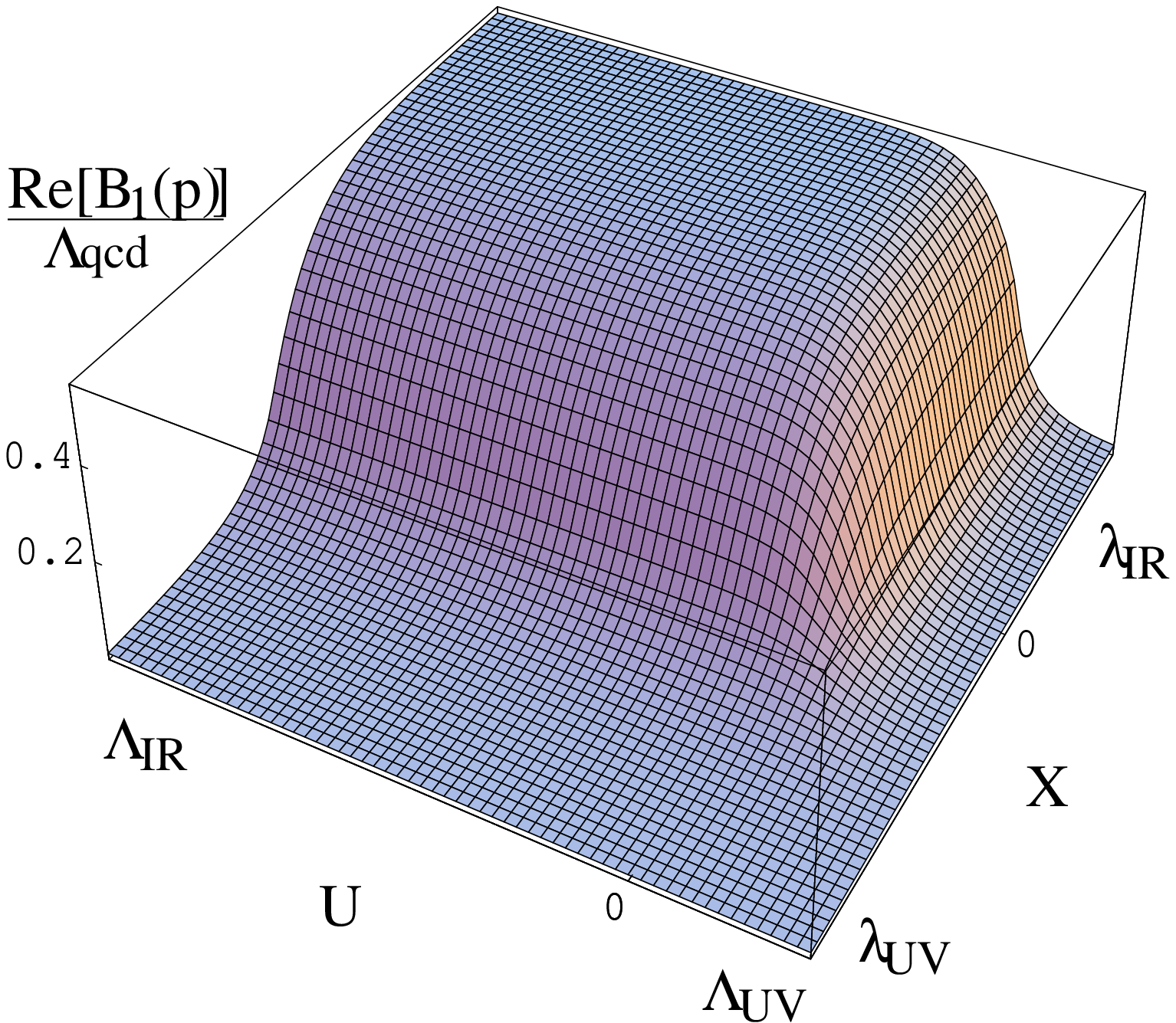}
\end{minipage}\hspace{1cm}
\begin{minipage}{6cm}
 \epsfxsize=6cm
 \epsfbox{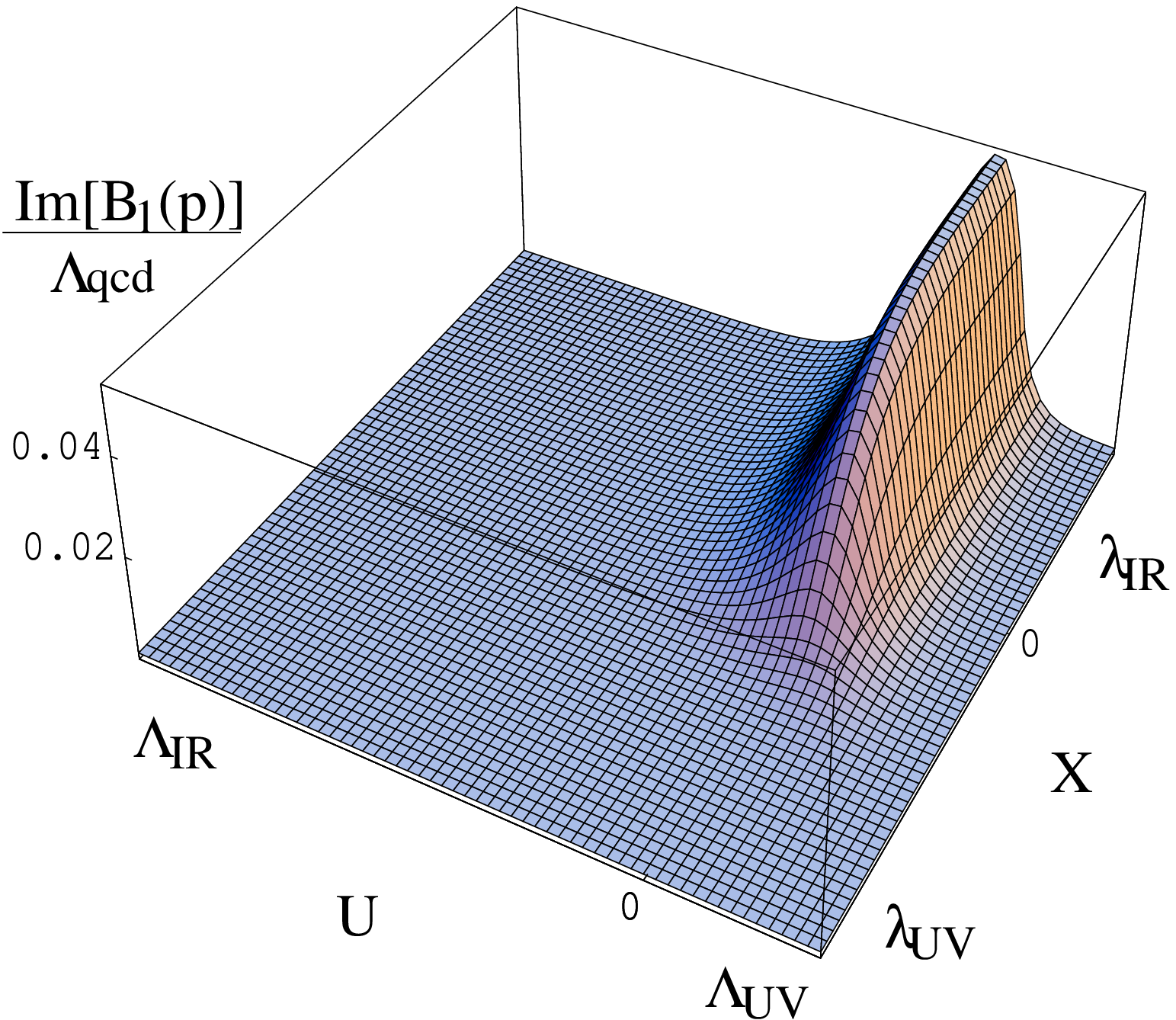} 
\end{minipage}

\vspace{0.3cm}

\begin{minipage}{12cm}
 Fig.~14$\cdot$ Momentum dependence of the Dirac mass function 
 $B_1(p)/\Lambda_{\rm qcd}$ at $\mu/\Lambda_{\rm qcd}=0.30$.
 Here $U = \ln (p_4/\Lambda_{\rm QCD} )$ and
 $X = \ln ( \bar{p}/\Lambda_{\rm QCD} )$.
\end{minipage}
\end{center}

\vspace{0.5cm}

We display the resultant values of 
$-\langle\bar\psi\psi\rangle^{1/3}_{\rm 1GeV}/{\Lambda_{\rm qcd}}$ and 
$\vert\bar{V}_{\rm sol}\vert^{1/4}/{\Lambda_{\rm qcd}}
 \cdot{\rm sgn[\bar{V}_{\rm sol}]}$ in Figs.~15 and~16,
 respectively. 
Considering Fig.~15 alone, it might seem that the phase transition
occurs around $\mu/\Lambda_{\rm qcd}=0.40$.(In Ref.\citen{Tani},
the chiral phase transition point was determined by the point where
their iteration converged to the trivial solution.) 
However, Fig.~16 shows that 
the value of $\bar{V}_{\rm sol}$ is 
positive for $0.36 < \mu/\Lambda_{\rm qcd} < 0.40$,
although it is negative for $\mu/\Lambda_{\rm qcd} < 0.36$.
In other words, the value of the effective potential
of the $\chi$SB vacuum for $0.36 < \mu/\Lambda_{\rm qcd} < 0.40$
is larger than that of the symmetric vacuum.
This implies that the $\chi$SB vacuum 
for $0.36 < \mu/\Lambda_{\rm qcd} < 0.40$
is a false vacuum,
and that the nontrivial solutions there
correspond to metastable states.
The existence of metastable states was found in Ref.~\citen{Bar}
by assuming the momentum dependence of the mass function,
and it was also shown in Ref.~\citen{Ha} 
by fully solving the SDE.
The value of the critical chemical potential
 $\mu_c/\Lambda_{\rm qcd} \simeq 0.36$ in
the present analysis is smaller than that obtained in
Ref.~\citen{Ha}, where the screening mass of the gluon was not
included and a slightly different form of the running coupling was
used.

\vspace{0.5cm}

\begin{center}
\begin{minipage}{12cm}
 \epsfxsize=12cm
 \epsfbox{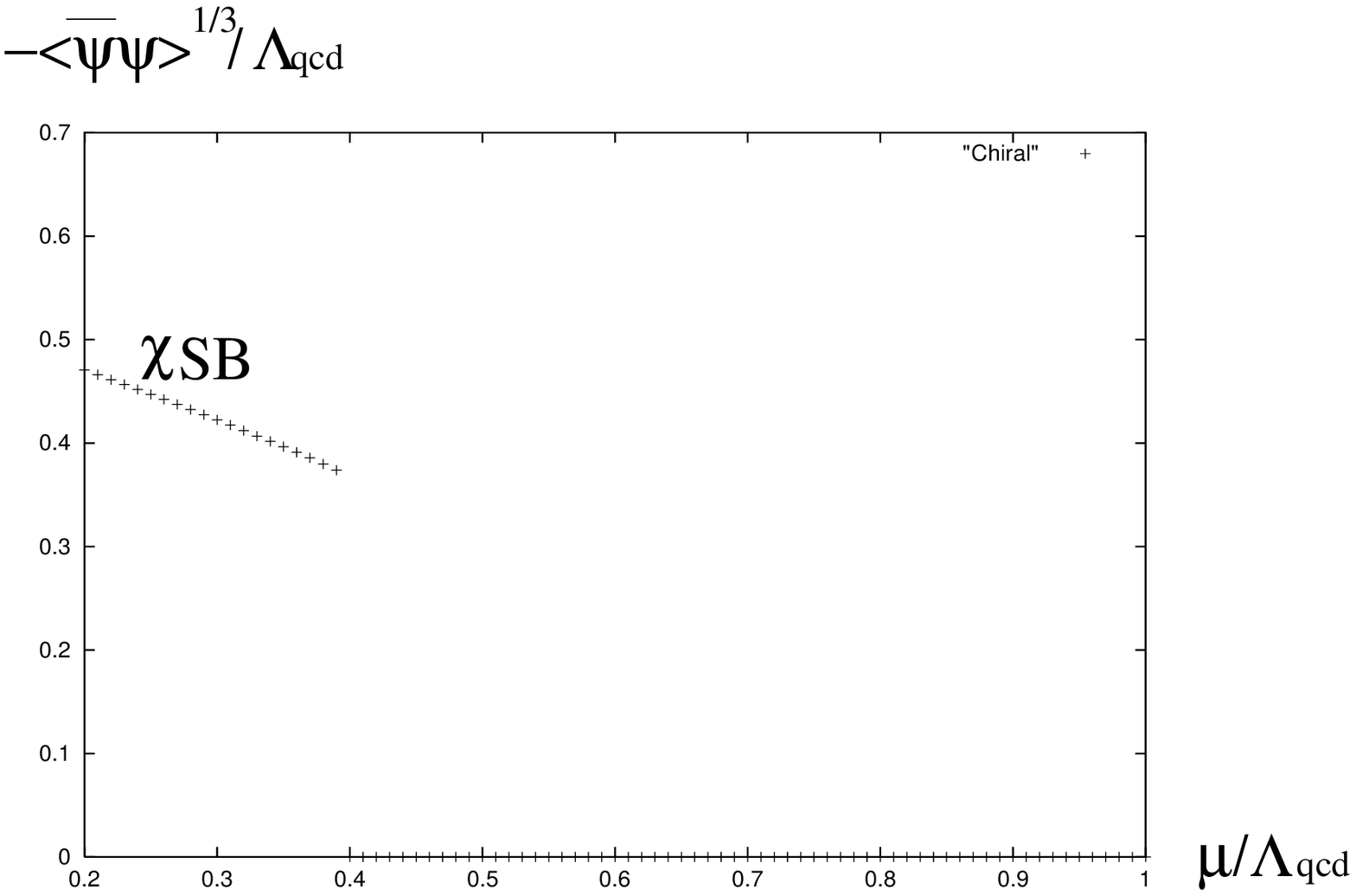}
 Fig.~15$\cdot$ Chemical potential dependence
 of $\langle\bar\psi\psi\rangle_{\rm 1GeV}/{\Lambda_{\rm qcd}^3}$ for
 $0.2\leq\mu/\Lambda_{\rm qcd}\leq{1.0}$.
\end{minipage}
\end{center}

\vspace{1cm}

\begin{center}
\begin{minipage}{12cm}
 \epsfxsize=12cm
 \epsfbox{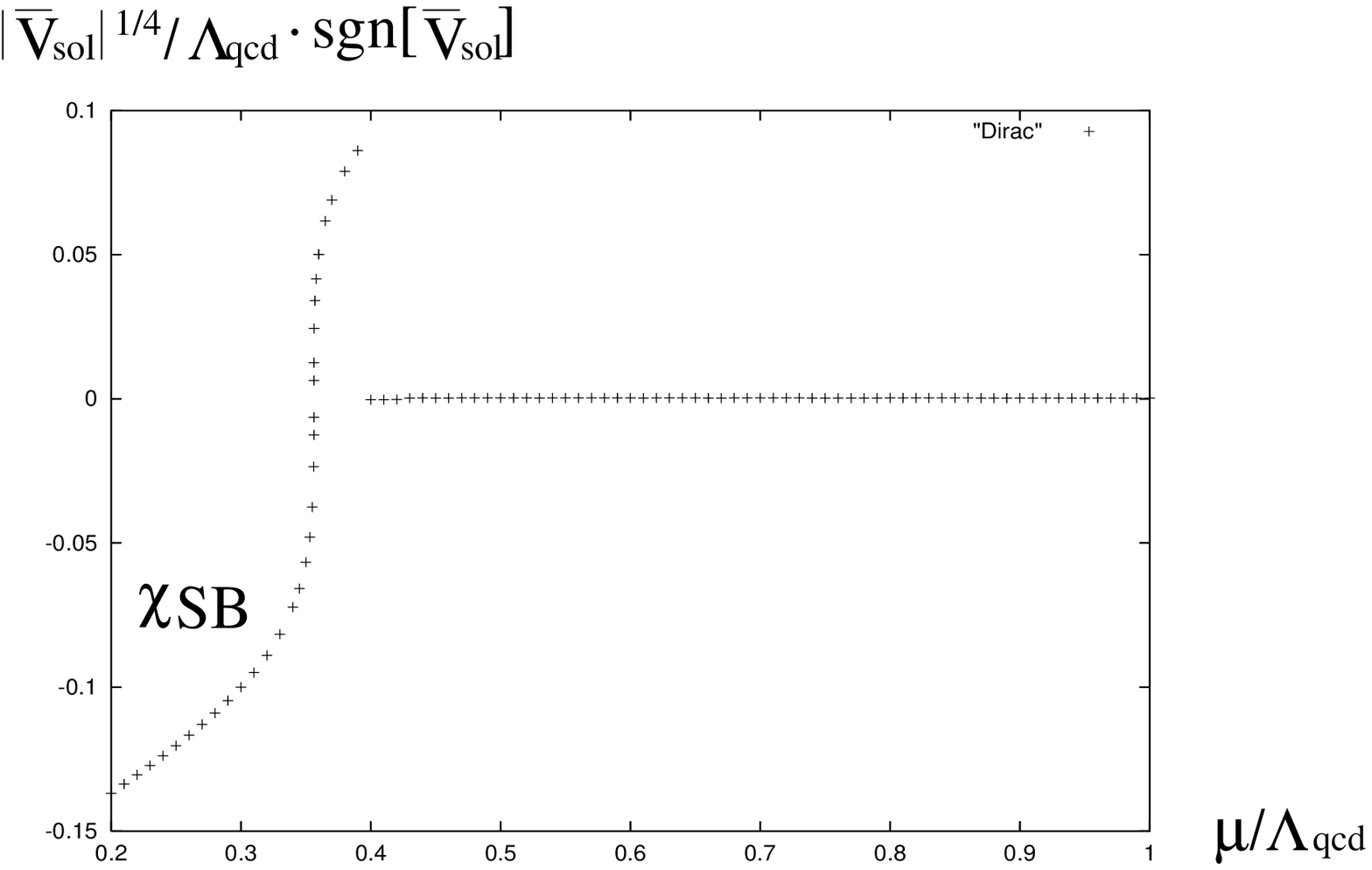}
 Fig.~16$\cdot$ Chemical potential dependence
 of $\vert\bar{V}_{\rm sol}\vert^{1/4}/{\Lambda_{\rm qcd}}
 \cdot{\rm sgn[\bar{V}_{\rm sol}]}$ for
 $0.2\leq\mu/\Lambda_{\rm qcd}\leq{1.0}$.
\end{minipage}
\end{center}

\vspace{0.5cm}

\subsection{Phase transition from hadronic phase to superconducting phase}
\label{Phase transition}
In this subsection we compare the values of the effective potential for 
the CSB solution with that for the $\chi$SB solution to determine
 the true vacuum.
We combine the effective potential for the CSB solution 
in Fig.~{10} with that for the $\chi$SB solution in Fig.~{16}
and display them in Fig.~{17}.
Figure~{17} shows that,
although the CSB vacuum is more stable than the trivial vacuum, 
the $\chi$SB vacuum is the most stable among these vacua in the low density
region. Hence the true vacuum in the low density region is the $\chi$SB vacuum.
This is natural, because the strength of the attractive force between two
quarks in the $\bar{3}$ channel is weaker than that between a quark and
antiquark in the singlet channel. 
We find that the chiral phase transition 
and the color superconducting phase transition occur 
simultaneously at $\mu_c/\Lambda_{\rm qcd}=0.344$
and that the phase transition is of first order.
We note that this phase transition occurs at lower density
than the chiral phase transition in the absence of
 color superconductivity.

To compare the diquark condensate with the chiral condensate,
we display them together in Fig.~{18}.
Note that these are scaled to $1$\,GeV by the renormalization group
formulas given in Eqs.~(\ref{bqqrenom}) and (\ref{qqrenom}). 
This figure shows that the diquark condensate for all $\mu$ 
is of the same order as the chiral condensate in the low density
region. The resultant values of these condensates 
at the critical chemical potential are
$\langle \bar{\psi} \psi \rangle_{\rm 1GeV} = 
- \left( 241\,\mbox{MeV} \right)^3$ and 
$\langle \psi \psi \rangle_{\rm 1GeV} =
- \left( 166\,\mbox{MeV} \right)^3$.

\vspace{1cm}

\begin{center}
\begin{minipage}{12cm}
 \epsfxsize=12cm
 \epsfbox{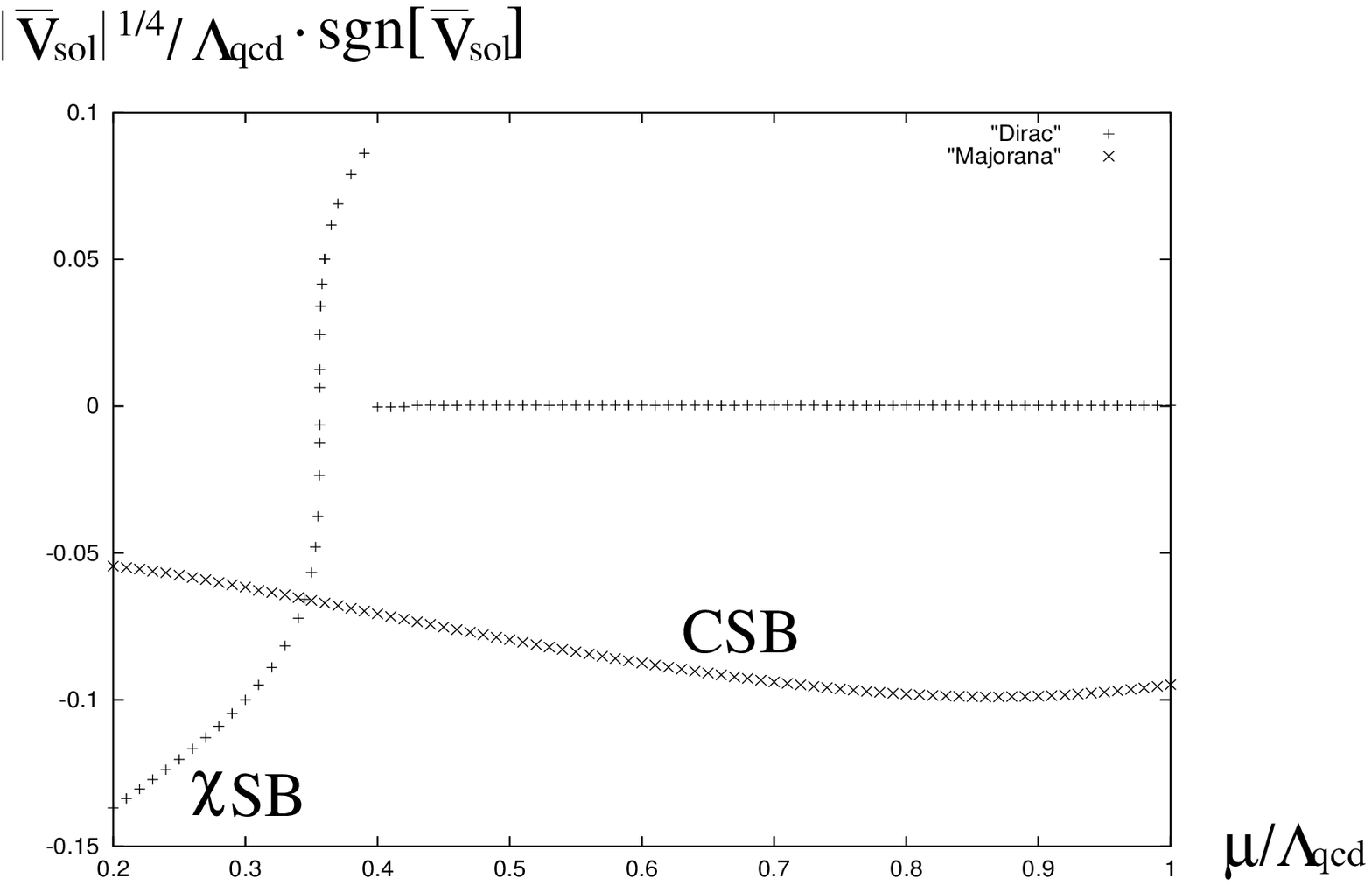}
 Fig.~17$\cdot$ Values of 
 $\vert\bar{V}_{\rm sol}\vert^{1/4}/{\Lambda_{\rm qcd}}
 \cdot{\rm sgn[\bar{V}_{\rm sol}]}$ ($0.2\leq\mu/\Lambda_{\rm qcd}\leq{1.0}$).
 Here ``$\chi$SB'' indicates that the inputs $B\neq{0}$ and 
 $\Delta=0$ were used, while ``CSB'' indicates that we used $B=0$ and 
 $\Delta\neq{0}$ as the initial trial functions.
\end{minipage}
\end{center}

\vspace{1cm}

\begin{center}
\begin{minipage}{12cm}
\epsfxsize=12cm
\epsfbox{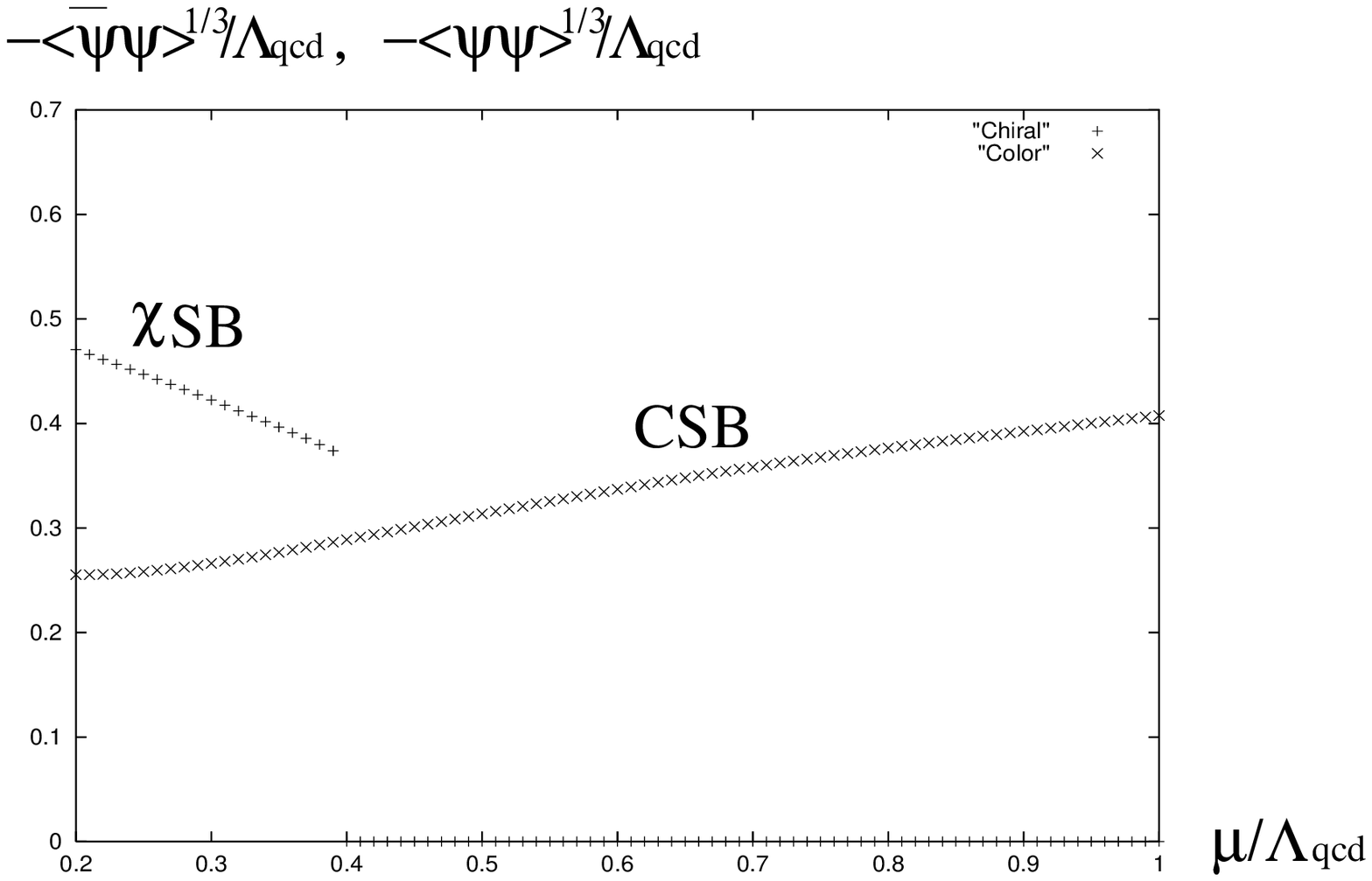}
 Fig.~18$\cdot$ Values of $-\langle\bar\psi\psi\rangle^{1/3}_{\rm 1GeV}
 /{\Lambda_{\rm qcd}}$
 and $-\langle\psi\psi\rangle^{1/3}_{\rm 1GeV}/{\Lambda_{\rm qcd}}$
 ($0.2\leq\mu/\Lambda_{\rm qcd}\leq{1.0}$).
Here ``$\chi$SB'' indicates the chiral condensate and
 ``CSB'' indicates the diquark condensate.
\end{minipage}
\end{center}

\vspace{0.5cm}

Finally, let us check the $t_f$ dependence of the critical chemical
potential. We list the values of the critical chemical potential $\mu_c$
for $t_f$=0.3--{0.7} in Table~I, together with the values of the
chiral condensate and the diquark condensate at $\mu_c$.
This table shows that the value of $\mu_c$ is almost independent of
 the value of $t_f$. Although the value of
 $\langle\bar{\psi}\psi\rangle$ depends slightly on that of $t_f$,
$\langle\psi\psi\rangle$ is quite stable with respect to the changes in $t_f$.

\vspace{0.5cm}

\begin{center}
\begin{minipage}{12cm}
 Table~I$\cdot$ Dependence of the critical chemical potential $\mu_c$ on the
 infrared cutoff parameter $t_f$. $\Lambda_{\rm qcd}$ is the value
 determined from $f_\pi=88$ MeV through Eq.~(\ref{psf}).
 The values of the chiral condensate and the diquark condensate 
 shown together are at the critical chemical potential $\mu_c$.
\[ 
 \begin{array}{ccccc}
  t_f & \Lambda_{\rm qcd}[{\rm MeV}] & {\mu}_c[{\rm MeV}] & -\langle\bar\psi\psi\rangle_{\rm 1GeV}^{1/3}[{\rm MeV}] & -\langle\psi\psi\rangle_{\rm 1GeV}^{1/3}[{\rm MeV}] \\ \cline{1-5}
   0.3 & 568 & 210 & 223 & 169 \\ \cline{1-5}
   0.4 & 579 & 207 & 231 & 168 \\ \cline{1-5}
   0.5 & 604 & 208 & 241 & 166 \\ \cline{1-5}
   0.6 & 647 & 214 & 254 & 167 \\ \cline{1-5} 
   0.7 & 712 & 223 & 269 & 168 \\ \cline{1-5} 
 \end{array}
\]
\end{minipage}
\end{center}

\vspace{0.5cm}

\section{Summary and discussion}
\label{Summary and Discussion}
We studied the phase structure in two-flavor dense QCD by solving the 
Schwinger-Dyson equations for the Dirac and Majorana masses
of the quark propagator with the improved ladder approximation in the
Landau gauge.
We showed that the momentum dependences of the Majorana masses of the quark
and antiquark are quite different, although the masses are of the same order.
This implies that the inclusion of two masses is important for studying
the intermediate density region.

The values of the chiral condensate (quark-antiquark condensate)
 and the diquark condensate were calculated using standard formulas.
We identified the condensates,
which were calculated with the cutoffs $\Lambda$,
with those renormalized at the scale $\Lambda$ in QCD, and 
then scaled them to the condensates at 1\,GeV 
using the leading renormalization group formulas.
The resultant value of the diquark condensate
is on the order of $-(200\,\mbox{MeV})^3$,
which is comparable to that of the chiral condensate for $\mu=0$.

The true vacuum was determined by comparing the values of the
effective potential at the solution.
We found that there exists a color symmetry breaking (CSB) solution 
for all the values of the chemical potential we studied, 
$0.2 \le \mu/\Lambda_{\rm qcd} \le 1$,
and that the CSB vacuum is more stable than the trivial vacuum in all regions.
Comparing the value of the effective potential for the CSB vacuum with
that for the chiral symmetry breaking ($\chi$SB) vacuum,
we showed that the $\chi$SB vacuum is more stable than the CSB vacuum
in the low density region.
We also found that the phase transition from the $\chi$SB vacuum to the CSB
vacuum is of first order. The critical chemical potential was 
determined as $\mu_c = 207$--$223$\,MeV.

Finally, we make several comments.
In an analysis employing the four-Fermi model,\cite{Shuryak}
it was apparently shown that there is a mixed phase,
in which both the chiral condensate and the diquark condensate exist,
in a small region of the chemical potential.
In our analysis, we tried to find the solution corresponding to this 
mixed phase.
However, we could not find such a solution in the present iteration
method, although we used several initial trial functions.
The value of the critical chemical potential obtained in the present
analysis may depend on
the detailed form of the running coupling. 
It is interesting to perform the similar analysis 
by using the different forms of the running coupling such
as the one used in Ref.\citen{Ha}.

In the present analysis we included the effects of the Debye mass
 and the Landau damping into the gluon propagator using 
the hard dense loop approximation,~\cite{Hong}
assuming smooth extrapolation from the high density region.
As we can see easily from the expression in Eq.~(\ref{gluonp}),
the magnetic mode explicitly breaks the Lorentz invariance, 
and therefore we cannot apply the form in Eq.~(\ref{gluonp})
 to the zero chemical potential case. 
 It may be interesting to compare the present results with
 those obtained using the gluon propagator without effect of Landau Damping.
This will be done elsewhere.

\section*{Acknowledgements}
This work is supported in part by a Grant-in-Aid for Scientific Research
[(A)\#12740144] (M. H.)

\appendix
\section{Quark Propagator and Schwinger-Dyson Equation}
\label{Quark Propagator and the Schwinger-Dyson Equation}
The explicit forms of the Nambu-Gorkov components of the full quark
 propagator in Eq.~(\ref{quarkp}) are given by

\begin{eqnarray}
 \label{SF11}
 S_{F11}(p)^{ab}_{ij}&=&
  S_{F22}(p)^{ab}_{ij}\vert[\mu\rightarrow-\mu;B(p)\rightarrow{B(-p)};
  \Delta^\pm(p)\rightarrow-\Delta^\mp(p)] \nonumber\\
  &=&i{R_+(p)^{-1}}^{ab}_{ij} \nonumber\\
 &=&\frac{i}{F(p,B_1,\Delta)}\nonumber\\
 &&\biggl[
  \biggl\{\biggl((p_0-\bar{p}+\mu)[(p_0-\mu)^2-\bar{p}^2-\{B_1(-p)\}^2]
  -(p_0-\bar{p}-\mu)\vert\Delta^-(p)\vert^2\biggr)\gamma_0 \nonumber\\
 &&\hspace{0.5cm}
  -B_1(-p)\Delta^-(p)\Delta^+(p)+B_1(p)
  [(p_0-\mu)^2-\bar{p}^2-\{B_1(-p)\}^2]\biggr\}
  \Lambda_p^+ \nonumber\\
 &&+\biggl\{\biggl((p_0+\bar{p}+\mu)
  [(p_0-\mu)^2-\bar{p}^2-\{B_1(-p)\}^2]
  -(p_0+\bar{p}-\mu)\vert\Delta^+(p)\vert^2\biggr)\gamma_0 \nonumber\\
 &&\hspace{0.5cm}
  -B_1(-p)\Delta^+(p)\Delta^-(p)
  +B_1(p)[(p_0-\mu)^2-\bar{p}^2-\{B_1(-p)\}^2]\biggr\}
  \Lambda_p^-
  \biggr]\delta_{ij}\tilde\delta^{ab} \nonumber\\
  &&+\frac{i}{(p_0+\mu)^2-\bar{p}^2-\{B_3(p)\}^2}
   \biggl[\biggl\{(p_0-\bar{p}+\mu)\gamma_0+B_3(p)\biggr\}\Lambda_p^+
   \nonumber\\
  &&\hspace{4cm} 
   +\biggl\{(p_0+\bar{p}+\mu)\gamma_0+B_3(p)\biggr\}\Lambda_p^-\biggr]
   \delta_{ij}\delta^{a3}\delta^{b3} \ \nonumber\\
\end{eqnarray}
and
\begin{eqnarray}
 \label{SF12}
 S_{F12}(p)^{ab}_{ij}&=&
  S_{F21}(p)^{ab}_{ij}\vert[\mu\rightarrow-\mu;B(p)\rightarrow{B(-p)};
  \Delta^\pm(p)\rightarrow-\Delta^\mp(p)] \nonumber\\
 &=&-i\{(p_0+\mu)\gamma^0-\vec{\gamma}\cdot\vec{p}-B(p)\}^{-1}
  \Delta(p){R_-(p)^{-1}}^{ab}_{ij} \nonumber\\
 &=&\frac{-i}{F(p,B_1,\Delta)}\biggl[
  \biggl\{\biggl((p_0-\bar{p}+\mu)\Delta^+(p)B_1(-p)
  -(p_0-\bar{p}-\mu)\Delta^-(p)B_1(p)\biggr)\gamma_0 \nonumber\\
 &&\hspace{0.5cm}
  +\biggl(\vert{B_1(p)}\vert^2\Delta^+(p)-[(p_0)^2-(\bar{p}+\mu)^2
  -\vert\Delta^+(p)\vert^2]\Delta^-(p)\biggr)\biggr\}\Lambda_p^+ \nonumber\\
 &&\hspace{1.5cm}+\biggl\{\biggl((p_0+\bar{p}+\mu)\Delta^-(p)B_1(-p)
  -(p_0+\bar{p}-\mu)\Delta^+(p)B_1(p)\biggr)\gamma_0 \nonumber\\
 &&\hspace{0.5cm}
  +\biggl(\vert{B_1(p)}\vert^2\Delta^-(p)-[(p_0)^2-(\bar{p}-\mu)^2
  -\vert\Delta^-(p)\vert^2]\Delta^+(p)\biggr)\biggr\}\Lambda_p^-
  \biggr]\gamma_5\epsilon_{ij}\epsilon^{ab3} \ , \nonumber\\
\end{eqnarray}
where
\begin{eqnarray}
 F(p,B_1,\Delta)
  &=&[(p_0+\mu)^2-\bar{p}^2-\{B_1(p)\}^2]
  [(p_0-\mu)^2-\bar{p}^2-\{B_1(-p)\}^2] \nonumber\\
 &&\hspace{0.5cm}-[(p_0)^2-(\bar{p}-\mu)^2]\vert\Delta^+(p)\vert^2
  -[(p_0)^2-(\bar{p}+\mu)^2]\vert\Delta^-(p)\vert^2 \nonumber\\
  &&\hspace{1cm}+\vert\Delta^+(p)\vert^2\vert\Delta^-(p)\vert^2
  +2B_1(p)B_1(-p)\Delta^+(p)\Delta^-(p) \ . \nonumber\\
\end{eqnarray}

Substituting the expression for $S_{F11}$ in Eq.~(\ref{SF11})
into the SDEs for $B_1$ and $B_3$ in Eqs.~(\ref{SD11a}) and
(\ref{SD11b}), we obtain
\begin{eqnarray}
\label{SDm}
 B_1(p)&=&-2\pi\int^\infty_{-\infty}
  \frac{dq_4}{(2\pi)^4}\int{d\bar{q}}\bar{q}^2\alpha_s
  \nonumber\\&&\times{K}_0(q_4,p_4,\bar{q},\bar{p})\biggl[
  \frac{5}{6}\frac{F_{+}(q,B_1,\Delta)}{F(q,B_1,\Delta)}
  +\frac{1}{2}\frac{B_3(q)}
  {(iq_4+\mu)^2-\bar{q}^2-\{B_3(q)\}^2}\biggr] \quad , \nonumber\\
\end{eqnarray}
\begin{eqnarray}
\label{SDm3}
 B_3(p)&=&-2\pi\int^\infty_{-\infty}
  \frac{dq_4}{(2\pi)^4}\int{d\bar{q}}\bar{q}^2\alpha_s
  \nonumber\\&&\times{K}_0(q_4,p_4,\bar{q},\bar{p})\biggl[
  \frac{F_{+}(q,B_1,\Delta)}{F(q,B_1,\Delta)}
  +\frac{1}{3}\frac{B_3(q)}
  {(iq_4+\mu)^2-\bar{q}^2-\{B_3(q)\}^2}\biggr] \quad , \nonumber\\
\end{eqnarray}
where
\begin{eqnarray}
 F_{+}(q,B_1,\Delta)&=&B_1(q)[(iq_4-\mu)^2-\bar{q}^2-\{B_1(-q)\}^2]
  -B_1(-q)\Delta^+(q)\Delta^-(q) \quad , \nonumber\\
\end{eqnarray}
with $q_4=-iq_0$ and $p_4=-ip_0$. The integration kernel $K_0$ is given by
\begin{eqnarray}
 K_0(q_4,p_4,\bar{q},\bar{p})
  &=&-i\int{d\Omega}D_{\mu\nu}(q-p)
  \mbox{tr}(\gamma^\mu\Lambda_q^\pm\gamma^\nu) \nonumber\\
 &=&4\cdot\frac{2\pi}{3\bar{q}\bar{p}}\log\frac
  {\vert\bar{q}+\bar{p}\vert^3+\omega_l^3}
  {\vert\bar{q}-\bar{p}\vert^3+\omega_l^3}
  +2\cdot{(1+d)}\frac{\pi}{\bar{q}\bar{p}}\log\frac
  {\vert\bar{q}+\bar{p}\vert^2+\omega^2+2M_D^2}
  {\vert\bar{q}-\bar{p}\vert^2+\omega^2+2M_D^2} \quad , \nonumber\\
\end{eqnarray}
where
\begin{eqnarray}
 \omega_l^3=\frac{\pi}{2}M_D^2\omega \ ,\qquad \omega=\vert{q_4-p_4}\vert \ .
\end{eqnarray}
When we set $\Delta^-=\Delta^+=0$, we have 
\begin{eqnarray}
 \frac{F_+(q,B_1,\Delta=0)}{F(q,B_1,\Delta=0)}&=&
  \frac{B_1(q)}{(iq_4+\mu)^2-\bar{q}^2-\{B_1(q)\}^2} \ .
\end{eqnarray}
If we further set $B_1= B_3$, 
the two equations in Eqs.~(\ref{SDm}) and (\ref{SDm3})
become identified,
This implies that $B_1= B_3$ is actually a solution of the SDEs 
for $\Delta^-=\Delta^+=0$.

Next, substituting Eq.~(\ref{SF12}) into the SDEs for $\Delta^-$ and 
$\Delta^+$ in Eqs.~(\ref{SD12}) and (\ref{SD21}), we obtain
\begin{eqnarray}
\label{SDdn}
 \Delta^-(p)&=&2\pi\int^\infty_{-\infty}
  \frac{dq_4}{(2\pi)^4}\int{d\bar{q}}\bar{q}^2\alpha_s \nonumber\\
 &&\times\biggl[K_1(q_4,p_4,\bar{q},\bar{p})\cdot
  \frac{2}{3}\cdot\frac{G_{+}(q,B_1,\Delta)}{F(q,B_1,\Delta)} 
  +K_2(q_4,p_4,\bar{q},\bar{p})\cdot
  \frac{2}{3}\cdot\frac{G_{-}(q,B_1,\Delta)}{F(q,B_1,\Delta)}\biggr] \ ,
  \nonumber\\
\end{eqnarray}
\begin{eqnarray}
\label{SDdp}
  \Delta^+(p)&=&2\pi\int^\infty_{-\infty}
  \frac{dq_4}{(2\pi)^4}\int{d\bar{q}}\bar{q}^2\alpha_s \nonumber\\
 &&\times\biggl[K_1(q_4,p_4,\bar{q},\bar{p})\cdot
  \frac{2}{3}\cdot\frac{G_{-}(q,B_1,\Delta)}{F(q,B_1,\Delta)} 
  +K_2(q_4,p_4,\bar{q},\bar{p})\cdot
  \frac{2}{3}\cdot\frac{G_{+}(q,B_1,\Delta)}{F(q,B_1,\Delta)}\biggr] \ ,
  \nonumber\\
\end{eqnarray}
where
\begin{eqnarray}
 G_{+}(q,B_1,\Delta)&=&B_1(q)B_1(-q)\Delta^+(q)+[(q_4)^2+(\bar{q}+\mu)^2
  +\vert\Delta^+(q)\vert^2]\Delta^-(q) \ , \nonumber\\ \\
 G_{-}(q,B_1,\Delta)&=&B_1(q)B_1(-q)\Delta^-(q)+[(q_4)^2+(\bar{q}-\mu)^2
  +\vert\Delta^-(q)\vert^2]\Delta^+(q) \ . \nonumber\\
\end{eqnarray}
The integration kernels $K_1$ and $K_2$ are given by
\begin{eqnarray}
 \lefteqn{K_1(q_4,p_4,\bar{q},\bar{p})
  =-i\int{d\Omega}D_{\mu\nu}(q-p)
  \mbox{tr}(\Lambda_p^\pm\gamma^\mu\Lambda_q^\mp\gamma^\nu)} \nonumber\\
  &=&-\pi\frac{(\bar{q}^2-\bar{p}^2)^2+\omega_l^4}
 {\sqrt{3}\omega_l^2\bar{q^2}\bar{p}^2}
 \arctan\biggl(\frac{\sqrt{3}\omega_l\min(\bar{q},\bar{p})}
 {\omega_l^2+\vert\bar{q}^2-\bar{p}^2\vert-\omega_l\max(\bar{q},\bar{p})}
 \biggr) \nonumber\\
 && -\pi\frac{(\bar{q}^2-\bar{p}^2)^2-\omega_l^4}
 {3\omega_l^2\bar{q^2}\bar{p}^2}
 \ln\frac{\omega_l+\vert\bar{q}+\bar{p}\vert}
 {\omega_l+\vert\bar{q}-\bar{p}\vert} 
 +\pi\frac{(\bar{q}^2-\bar{p}^2)^2-\omega_l^4}
 {6\omega_l^2\bar{q^2}\bar{p}^2}
 \ln\frac{\omega_l^2+\vert\bar{q}+\bar{p}\vert^2-
 \omega_l\vert\bar{q}+\bar{p}\vert}
 {\omega_l^2+\vert\bar{q}-\bar{p}\vert^2-\omega_l\vert\bar{q}-\bar{p}\vert}
 \nonumber\\
 &&+\pi\frac{4}{3\bar{q}\bar{p}}
  \ln\frac{\omega_l^3+\vert\bar{q}+\bar{p}\vert^3}
  {\omega_l^3+\vert\bar{q}-\bar{p}\vert^3} 
  +\frac{\pi}{2\bar{q}^2\bar{p}^2}\biggl[
  \frac{(\bar{q}^2-\bar{p}^2)^2}{2M_D^2+\omega^2}
  \ln\frac{(\bar{q}+\bar{p})^2}{(\bar{q}-\bar{p})^2} \nonumber\\
 &&\hspace{0.5cm}+\frac{\{(\bar{q}+\bar{p})^2+2M_D^2+\omega^2\}
  \{(2M_D^2+\omega^2)^2+(\bar{q}-\bar{p})^2\omega^2\}}
  {2M_D^2(2M_D^2+\omega^2)}\ln\frac{(\bar{q}+\bar{p})^2+2M_D^2+\omega^2}
  {(\bar{q}-\bar{p})^2+2M_D^2+\omega^2} \nonumber\\ 
 &&\hspace{0.5cm}
  -\frac{\{(\bar{q}+\bar{p})^2+\omega^2\}\{(\bar{q}-\bar{p})^2+\omega^2\}}
  {2M_D^2}\ln\frac{(\bar{q}+\bar{p})^2+\omega^2}
  {(\bar{q}-\bar{p})^2+\omega^2}
  \biggr] \nonumber\\
 &&+\pi{d}\biggl[\frac{2}{\bar{q}\bar{p}}
  -\frac{(\bar{q}-\bar{p})^2+\omega^2}{2\bar{q}^2\bar{p}^2}
  \ln\frac{(\bar{q}+\bar{p})^2+\omega^2}
  {(\bar{q}-\bar{p})^2+\omega^2}\biggr] \ ,
  \nonumber\\
\end{eqnarray}
\begin{eqnarray}
 \lefteqn{K_2(q_4,p_4,\bar{q},\bar{p})
  =-i\int{d\Omega}D_{\mu\nu}(q-p)
  \mbox{tr}(\Lambda_p^\pm\gamma^\mu\Lambda_q^\pm\gamma^\nu)} \nonumber\\
 &=&\pi\frac{(\bar{q}^2-\bar{p}^2)^2+\omega_l^4}
  {\sqrt{3}\omega_l^2\bar{q^2}\bar{p}^2}
  \arctan\biggl(\frac{\sqrt{3}\omega_l\min(\bar{q},\bar{p})}
  {\omega_l^2+\vert\bar{q}^2-\bar{p}^2\vert-\omega_l\max(\bar{q},\bar{p})}
  \biggr) \nonumber\\
 && +\pi\frac{(\bar{q}^2-\bar{p}^2)^2-\omega_l^4}
  {3\omega_l^2\bar{q^2}\bar{p}^2}
  \ln\frac{\omega_l+\vert\bar{q}+\bar{p}\vert}
  {\omega_l+\vert\bar{q}-\bar{p}\vert}
  -\pi\frac{(\bar{q}^2-\bar{p}^2)^2-\omega_l^4}
  {6\omega_l^2\bar{q^2}\bar{p}^2}
  \ln\frac{\omega_l^2+\vert\bar{q}+\bar{p}\vert^2-
   \omega_l\vert\bar{q}+\bar{p}\vert}
  {\omega_l^2+\vert\bar{q}-\bar{p}\vert^2-\omega_l\vert\bar{q}-\bar{p}\vert}
  \nonumber\\
 &&+\pi\frac{4}{3\bar{q}\bar{p}}
  \ln\frac{\omega_l^3+\vert\bar{q}+\bar{p}\vert^3}
  {\omega_l^3+\vert\bar{q}-\bar{p}\vert^3} 
  -\frac{\pi}{2\bar{q}^2\bar{p}^2}\biggl[
  \frac{(\bar{q}^2-\bar{p}^2)^2}{2M_D^2+\omega^2}
  \ln\frac{(\bar{q}+\bar{p})^2}{(\bar{q}-\bar{p})^2} \nonumber\\
 &&\hspace{0.5cm}+\frac{\{(\bar{q}-\bar{p})^2+2M_D^2+\omega^2\}
  \{(2M_D^2+\omega^2)^2+(\bar{q}+\bar{p})^2\omega^2\}}
  {2M_D^2(2M_D^2+\omega^2)}\ln\frac{(\bar{q}+\bar{p})^2+2M_D^2+\omega^2}
  {(\bar{q}-\bar{p})^2+2M_D^2+\omega^2} \nonumber\\ 
 &&\hspace{0.5cm}
  -\frac{\{(\bar{q}+\bar{p})^2+\omega^2\}\{(\bar{q}-\bar{p})^2+\omega^2\}}
  {2M_D^2}\ln\frac{(\bar{q}+\bar{p})^2+\omega^2}
  {(\bar{q}-\bar{p})^2+\omega^2}
  \biggr] \nonumber\\
 &&-\pi{d}\biggl[\frac{2}{\bar{q}\bar{p}}
  -\frac{(\bar{q}+\bar{p})^2+\omega^2}{2\bar{q}^2\bar{p}^2}
  \ln\frac{(\bar{q}+\bar{p})^2+\omega^2}
  {(\bar{q}-\bar{p})^2+\omega^2}\biggr] \ .
  \nonumber\\
\end{eqnarray}
We note that for $B_1=0$ we have
\begin{eqnarray}
 \frac{G_{\pm}(q,B_1=0,\Delta)}{F(q,B_1=0,\Delta)}&=&
  \frac{-\Delta^\mp(q)}{(q_4)^2+(\bar{q}\mp\mu)^2+\vert\Delta^\mp(q)\vert^2}
  \ . 
\end{eqnarray}
Then the SDEs (\ref{SDdn}) and (\ref{SDdp}) are identical to the well known
forms given in,~e.g.,~Ref.~\citen{Hong}.

\section{Convenient Formulas}
\label{Convenient Formulae}
In this appendix
we give several convenient formulas to obtain the explicit forms of
the Schwinger-Dyson equations and the effective potential given in
\S\ref{Effective potential and Schwinger-Dyson equation} and
Appendix~\ref{Quark Propagator and the Schwinger-Dyson Equation}. 

As shown in \S\ref{Gluon propagator and the running couplinig},
 we use the Landau gauge in the present analysis,
and two polarization tensors are used in Eq.~(\ref{gluonp}).
When we use the general covariant gauge, as done in Ref.~\citen{Hong},
we need the three independent polarization tensors defined as

\begin{eqnarray}
 &&O^{(1)}_{\mu\nu}=P^\bot_{\mu\nu}+\frac{(u\cdot{k})^2}{(u\cdot{k})^2-k^2}
  P^u_{\mu\nu} \ , \quad 
  O^{(2)}_{\mu\nu}=-\frac{(u\cdot{k})^2}{(u\cdot{k})^2-k^2}P^u_{\mu\nu} \ ,
  \quad O^{(3)}_{\mu\nu}=\frac{k_\mu{k}_\nu}{k^2} \ , \nonumber\\
\end{eqnarray}
where
\begin{eqnarray}
 &&P^\bot_{\mu\nu}=g_{\mu\nu}-\frac{k_\mu{k}_\nu}{k^2} \ ,
  \hspace{1cm} 
  P^u_{\mu\nu}=\frac{k_\mu{k}_\nu}{k^2}-\frac{k_\mu{u}_\nu+u_\mu{k}_\nu}
  {u\cdot{k}}+\frac{u_\mu{u}_\nu}{(u\cdot{k})^2}k^2 \ .
\end{eqnarray}

The following formulas are convenient to obtain the SDEs:
\begin{eqnarray}
 O^{(1)}_{\mu\nu}\mbox{tr}(\gamma^\mu\Lambda_q^{\pm}\gamma^\nu)=4 \ , \qquad
  O^{(i)}_{\mu\nu}\mbox{tr}(\gamma^\mu\Lambda_q^{\pm}\gamma^\nu)=2 \ , \qquad 
  (i=2,3) \nonumber
\end{eqnarray}
\begin{eqnarray}
 O^{(1)}_{\mu\nu}\mbox{tr}
  (\Lambda_p^{\pm}\gamma^\mu\Lambda_q^{\pm}\gamma^\nu)
  &=&2(1+t)\frac{\bar{q}^2+\bar{p}^2-\bar{q}\bar{p}(1+t)}
  {\bar{q}^2+\bar{p}^2-2\bar{q}\bar{p}t} \ ,\nonumber\\
 O^{(1)}_{\mu\nu}\mbox{tr}
  (\Lambda_p^{\pm}\gamma^\mu\Lambda_q^{\mp}\gamma^\nu)
  &=&2(1-t)\frac{\bar{q}^2+\bar{p}^2+\bar{q}\bar{p}(1-t)}
  {\bar{q}^2+\bar{p}^2-2\bar{q}\bar{p}t} \ ,\nonumber\\
 O^{(2)}_{\mu\nu}\mbox{tr}
  (\Lambda_p^{\pm}\gamma^\mu\Lambda_q^{\pm}\gamma^\nu)
  &=&2(1-t)\frac{\bar{q}^2+\bar{p}^2+\bar{q}\bar{p}(1-t)}
  {\bar{q}^2+\bar{p}^2-2\bar{q}\bar{p}t}
  -(1-t)\frac{(\bar{q}+\bar{p})^2+(q_4-p_4)^2}
  {\bar{q}^2+\bar{p}^2-2\bar{q}\bar{p}t+(q_4-p_4)^2} \ ,\nonumber\\ 
 O^{(2)}_{\mu\nu}\mbox{tr}
  (\Lambda_p^{\pm}\gamma^\mu\Lambda_q^{\mp}\gamma^\nu)
  &=&2(1+t)\frac{\bar{q}^2+\bar{p}^2-\bar{q}\bar{p}(1+t)}
  {\bar{q}^2+\bar{p}^2-2\bar{q}\bar{p}t}
  -(1+t)\frac{(\bar{q}-\bar{p})^2+(q_4-p_4)^2}
  {\bar{q}^2+\bar{p}^2-2\bar{q}\bar{p}t+(q_4-p_4)^2} \ ,\nonumber\\ 
 O^{(3)}_{\mu\nu}\mbox{tr}
  (\Lambda_p^{\pm}\gamma^\mu\Lambda_q^{\pm}\gamma^\nu)
  &=&(1-t)\frac{(\bar{q}+\bar{p})^2+(q_4-p_4)^2}
  {\bar{q}^2+\bar{p}^2-2\bar{q}\bar{p}t+(q_4-p_4)^2} \ ,\nonumber\\
 O^{(3)}_{\mu\nu}\mbox{tr}
  (\Lambda_p^{\pm}\gamma^\mu\Lambda_q^{\mp}\gamma^\nu)
  &=&(1+t)\frac{(\bar{q}-\bar{p})^2+(q_4-p_4)^2}
  {\bar{q}^2+\bar{p}^2-2\bar{q}\bar{p}t+(q_4-p_4)^2} \ , \nonumber\\ 
\end{eqnarray}
where $t=\vec{p}\cdot\vec{q}/\vert{p}\vert\vert{q}\vert$. 
Angle integrations in the SDEs are performed using the 
following formulas:\cite{Hong}
\begin{eqnarray}
 &&\int\frac{d\Omega\vert\vec{q}-\vec{p}\vert}
  {\vert\vec{q}-\vec{p}\vert^3+\omega_l^3}
  =\frac{2\pi}{3\bar{q}\bar{p}}\log\frac
  {\vert\bar{q}+\bar{p}\vert^3+\omega_l^3}
  {\vert\bar{q}-\bar{p}\vert^3+\omega_l^3} \ , \hspace{10cm}\nonumber\\
 &&\int\frac{d\Omega}{\vert\vec{q}-\vec{p}\vert^2+\omega^2+2M_D^2}
  =\frac{\pi}{\bar{q}\bar{p}}\log\frac
  {\vert\bar{q}+\bar{p}\vert^2+\omega^2+2M_D^2}
  {\vert\bar{q}-\bar{p}\vert^2+\omega^2+2M_D^2} \ , \nonumber 
\end{eqnarray}
\begin{eqnarray}
 \lefteqn{\int\frac{d\Omega\vert\vec{q}-\vec{p}\vert}
  {\vert\vec{q}-\vec{p}\vert^3+\omega_l^3}O^{(1)}_{\mu\nu}
  \mbox{tr}(\Lambda_p^{\pm}\gamma^\mu\Lambda_q^{\pm}\gamma^\nu)}\nonumber\\
  &&=\pi\biggl[-\frac{2}{\bar{q}\bar{p}}
  -\frac{(\bar{q}^2-\bar{p}^2)^2+\omega_l^4}
  {\sqrt{3}\omega_l^2\bar{q^2}\bar{p}^2}
 \arctan\biggl(\frac{\sqrt{3}\omega_l\min(\bar{q},\bar{p})}
 {\omega_l^2+\vert\bar{q}^2-\bar{p}^2\vert-\omega_l\max(\bar{q},\bar{p})}
 \biggr) \nonumber\\
 &&\hspace{0.5cm}+\frac{(\bar{q}^2-\bar{p}^2)^2-\omega_l^4}
 {3\omega_l^2\bar{q^2}\bar{p}^2}
 \ln\frac{\omega_l+\vert\bar{q}+\bar{p}\vert}
 {\omega_l+\vert\bar{q}-\bar{p}\vert}
 -\frac{(\bar{q}^2-\bar{p}^2)^2-\omega_l^4}
 {6\omega_l^2\bar{q^2}\bar{p}^2}
 \ln\frac{\omega_l^2+\vert\bar{q}+\bar{p}\vert^2-
 \omega_l\vert\bar{q}+\bar{p}\vert}
 {\omega_l^2+\vert\bar{q}-\bar{p}\vert^2-\omega_l\vert\bar{q}-\bar{p}\vert}
 \nonumber\\
 &&\hspace{0.5cm}+\frac{4}{3\bar{q}\bar{p}}
  \ln\frac{\omega_l^3+\vert\bar{q}+\bar{p}\vert^3}
 {\omega_l^3+\vert\bar{q}-\bar{p}\vert^3}\biggr] \ , \nonumber
\end{eqnarray}
\begin{eqnarray}
 \lefteqn{\int\frac{d\Omega\vert\vec{q}-\vec{p}\vert}
  {\vert\vec{q}-\vec{p}\vert^3+\omega_l^3}O^{(1)}_{\mu\nu}
  \mbox{tr}(\Lambda_p^{\pm}\gamma^\mu\Lambda_q^{\mp}\gamma^\nu)}\nonumber\\ 
 &&=\pi\biggl[\frac{2}{\bar{q}\bar{p}}
  -\frac{(\bar{q}^2-\bar{p}^2)^2+\omega_l^4}
  {\sqrt{3}\omega_l^2\bar{q^2}\bar{p}^2}
  \arctan\biggl(\frac{\sqrt{3}\omega_l\min(\bar{q},\bar{p})}
  {\omega_l^2+\vert\bar{q}^2-\bar{p}^2\vert-\omega_l\max(\bar{q},\bar{p})}
  \biggr) \nonumber\\
 &&\hspace{0.5cm}-\frac{(\bar{q}^2-\bar{p}^2)^2-\omega_l^4}
  {3\omega_l^2\bar{q^2}\bar{p}^2}
  \ln\frac{\omega_l+\vert\bar{q}+\bar{p}\vert}
  {\omega_l+\vert\bar{q}-\bar{p}\vert}
  +\frac{(\bar{q}^2-\bar{p}^2)^2-\omega_l^4}
  {6\omega_l^2\bar{q^2}\bar{p}^2}
  \ln\frac{\omega_l^2+\vert\bar{q}+\bar{p}\vert^2-
  \omega_l\vert\bar{q}+\bar{p}\vert}
  {\omega_l^2+\vert\bar{q}-\bar{p}\vert^2-\omega_l\vert\bar{q}-\bar{p}\vert}
  \nonumber\\
 &&\hspace{0.5cm}+\frac{4}{3\bar{q}\bar{p}}
  \ln\frac{\omega_l^3+\vert\bar{q}+\bar{p}\vert^3}
  {\omega_l^3+\vert\bar{q}-\bar{p}\vert^3}\biggr] \ ,\nonumber
\end{eqnarray}
\begin{eqnarray}
 \lefteqn{\int\frac{d\Omega}{\vert\vec{q}-\vec{p}\vert^2+\omega^2+2M_D^2}
  O^{(2)}_{\mu\nu}\mbox{tr}(\Lambda_p^{\pm}\gamma^\mu
  \Lambda_q^{\pm}\gamma^\nu)} \nonumber\\ 
 &&=\frac{2\pi}{\bar{q}\bar{p}}+\frac{\pi}{2\bar{q}^2\bar{p}^2}\biggl[
  -\frac{(\bar{q}^2-\bar{p}^2)^2}{2M_D^2+\omega^2}
  \ln\frac{(\bar{q}+\bar{p})^2}{(\bar{q}-\bar{p})^2} \nonumber\\
 &&\hspace{0.5cm}-\frac{\{(\bar{q}-\bar{p})^2+2M_D^2+\omega^2\}
  \{(2M_D^2+\omega^2)^2+(\bar{q}+\bar{p})^2\omega^2\}}
  {2M_D^2(2M_D^2+\omega^2)}
  \ln\frac{(\bar{q}+\bar{p})^2+2M_D^2+\omega^2}
  {(\bar{q}-\bar{p})^2+2M_D^2+\omega^2} \nonumber\\ 
 &&\hspace{0.5cm}+\frac{\{(\bar{q}+\bar{p})^2+\omega^2\}
  \{(\bar{q}-\bar{p})^2+\omega^2\}}
  {2M_D^2}\ln\frac{(\bar{q}+\bar{p})^2+\omega^2}
  {(\bar{q}-\bar{p})^2+\omega^2}
  \biggr] \ , \nonumber
\end{eqnarray}
\begin{eqnarray}
 \lefteqn{\int\frac{d\Omega}{\vert\vec{q}-\vec{p}\vert^2+\omega^2+2M_D^2}
  O^{(2)}_{\mu\nu}\mbox{tr}(\Lambda_p^{\pm}\gamma^\mu
  \Lambda_q^{\mp}\gamma^\nu)}\nonumber\\ 
 &&=-\frac{2\pi}{\bar{q}\bar{p}}+\frac{\pi}{2\bar{q}^2\bar{p}^2}\biggl[
  \frac{(\bar{q}^2-\bar{p}^2)^2}{2M_D^2+\omega^2}
  \ln\frac{(\bar{q}+\bar{p})^2}{(\bar{q}-\bar{p})^2} \nonumber\\
 &&\hspace{0.5cm}+\frac{\{(\bar{q}+\bar{p})^2+2M_D^2+\omega^2\}
  \{(2M_D^2+\omega^2)^2+(\bar{q}-\bar{p})^2\omega^2\}}
  {2M_D^2(2M_D^2+\omega^2)}\ln\frac{(\bar{q}+\bar{p})^2+2M_D^2+\omega^2}
  {(\bar{q}-\bar{p})^2+2M_D^2+\omega^2} \nonumber\\ 
 &&\hspace{0.5cm}-\frac{\{(\bar{q}+\bar{p})^2+\omega^2\}
  \{(\bar{q}-\bar{p})^2+\omega^2\}}
  {2M_D^2}\ln\frac{(\bar{q}+\bar{p})^2+\omega^2}
  {(\bar{q}-\bar{p})^2+\omega^2}\biggr] \ , \nonumber
\end{eqnarray}
\begin{eqnarray}
  \int\frac{d\Omega}{\vert\vec{q}-\vec{p}\vert^2+\omega^2}O^{(3)}_{\mu\nu}
  \mbox{tr}(\Lambda_p^{\pm}\gamma^\mu\Lambda_q^{\pm}\gamma^\nu) 
  &=&\pi\biggl[-\frac{2}{\bar{q}\bar{p}}
  +\frac{(\bar{q}-\bar{p})^2+\omega^2}{2\bar{q}^2\bar{p}^2}
  \ln\frac{(\bar{q}+\bar{p})^2+\omega^2}
  {(\bar{q}-\bar{p})^2+\omega^2}\biggr],\nonumber\\
  \int\frac{d\Omega}{\vert\vec{q}-\vec{p}\vert^2+\omega^2}
  O^{(3)}_{\mu\nu}\mbox{tr}(\Lambda_p^{\pm}\gamma^\mu
  \Lambda_q^{\mp}\gamma^\nu)
  &=&\pi\biggl[\frac{2}{\bar{q}\bar{p}}
  -\frac{(\bar{q}-\bar{p})^2+\omega^2}{2\bar{q}^2\bar{p}^2}
  \ln\frac{(\bar{q}+\bar{p})^2+\omega^2}
  {(\bar{q}-\bar{p})^2+\omega^2}\biggr] \ . \nonumber\\
\end{eqnarray}

Here we list the products of the color matrices in the SDE.
The completeness relation leads to  
\begin{eqnarray}
 \sum_{A=1}^8(T^A)_{a^{\prime}a}(T^A)_{b^{\prime}b}
  &=&\frac{1}{2}\delta_{a^{\prime}b}\delta_{ab^{\prime}}
  -\frac{1}{6}\delta_{aa^{\prime}}\delta_{bb^{\prime}} \ . 
\end{eqnarray}
By using the above formula the products of the color matrices in 
the SDEs (\ref{SD11a}) and (\ref{SD11b}) can be written
\begin{eqnarray}
 \sum_{A=1}^8(T^A)_{aa^{\prime}}\tilde\delta^{a^{\prime}b^{\prime}}
  (T^A)_{b^{\prime}b}
 &=&\frac{5}{6}\tilde\delta_{ab}+\delta_{a3}\delta_{b3} \ , \nonumber\\
 \sum_{A=1}^8(T^A)_{aa^{\prime}}\delta_{a^{\prime}3}\delta_{b^{\prime}3}
  (T^A)_{b^{\prime}b}
 &=&\frac{1}{2}\tilde\delta_{ab}+\frac{1}{3}\delta_{a3}\delta_{b3} \ . 
\nonumber\\
\end{eqnarray}
The product in the SDEs (\ref{SD12}) and (\ref{SD21}) can be written   
\begin{eqnarray}
 \sum_{A=1}^8(T^A)_{aa^{\prime}}\epsilon^{a^{\prime}b^{\prime}3}
  (T^A)_{bb^{\prime}}
 &=&-\frac{2}{3}\epsilon^{ab3} \ . 
\end{eqnarray}

To obtain the effective potential we need to take the trace and
 the determinant over the color, flavor, spinor and
Nambu-Gorkov indices. The determinant in the first term of 
 Eq.~(\ref{Efa}) is given by
\begin{eqnarray}
 \mbox{Det}\{{i}S_F(p)^{-1}\}
  &=&\{F(p,B_1,\Delta)\}^8\{F(p,B_3,0)\}^4 \ .
\end{eqnarray}
The trace in the second term becomes
\begin{eqnarray}
 \mbox{Tr}\{S_{F}^{(0)}(p)^{-1}S_{F}(p)\} 
  &=&8\biggl[\frac{2}{F(p,B_1,\Delta)}\biggl\{
  2(p_+^2-\bar{p}^2)(p_-^2-\bar{p}^2) \nonumber\\
 &&\hspace{0.5cm}-(p_+^2-\bar{p}^2)\{B_1(-p)\}^2-(p_-^2-\bar{p}^2)\{B_1(p)\}^2
  \nonumber\\
 &&\hspace{0.3cm}-\biggl((p_0)^2-(\bar{p}+\mu)^2\biggr)\vert\Delta^-\vert^2
  -\biggl((p_0)^2-(\bar{p}-\mu)^2\biggr)\vert\Delta^+\vert^2
  \biggr\} \nonumber\\
 &&+\frac{(p_0+\mu)^2-\bar{p}^2}
  {(p_0+\mu)^2-\bar{p}^2-\{B_3(p)\}^2}
  +\frac{(p_0-\mu)^2-\bar{p}^2}
  {(p_0-\mu)^2-\bar{p}^2-\{B_3(-p)\}^2}\biggr] \ . \nonumber\\
\end{eqnarray}

\end{document}